\documentclass[twocolumn,superscriptaddress,showpacs,preprintnumbers,amsmath,amssymb]{revtex4}

\usepackage{bm}
\usepackage{amssymb}
\usepackage{amsmath}
\usepackage{graphicx}
\usepackage{booktabs}
\usepackage{longtable}
\usepackage{tabularx}
\usepackage{array}
\usepackage{tabu}
\begin{document}

\title{Spin-orbit splittings of neutron states in $N = 20$ isotones from covariant density functionals and their extensions}

\author{Konstantinos Karakatsanis}
\email[]{kokaraka@auth.gr}
\affiliation{Physics Department,
	Aristotle University of Thessaloniki,
	Thessaloniki GR-54124, Greece}
\author{G. A. Lalazissis}
\affiliation{Physics Department,
	Aristotle University of Thessaloniki,
	Thessaloniki GR-54124, Greece}
\author{Peter Ring}
\affiliation{Physik Department, Technische Universit\"at Munchen, D-85747 Garching, Germany}
\author{Elena Litvinova}
\affiliation{Department of Physics, Western Michigan University, Kalamazoo MI 49008-5252 USA}
\affiliation{National
Superconducting Cyclotron Laboratory, Michigan State University,
East Lansing, MI 48824-1321, USA}

\date{\today}

\begin{abstract}
Spin-orbit splitting is an essential ingredient for our understanding of the shell structure in nuclei.
One of the most important advantages of relativistic mean-field (RMF) models in nuclear physics is the fact that the large spin-orbit (SO) potential emerges automatically from the inclusion of Lorentz-scalar and -vector potentials in the Dirac equation. It is therefore of great importance to compare the results of such models with experimental data. We investigate the size of $2p$- and $1f$ splittings for the isotone chain $^{40}$Ca, $^{38}$Ar, $^{36}$S, and $^{34}$Si in the framework of various relativistic and nonrelativistic density functionals. They are compared with the results of nonrelativistic models and with recent experimental data.
\end{abstract}

\pacs{}
\maketitle

\section{\label{introduction} Introduction}

Self-consistent mean-field models in the framework of nuclear density functional theory provide a very successful way to study nuclear structure phenomena throughout the entire nuclear chart. The nucleons are treated as independent particles moving inside the nucleus under the influence of various potentials, derived from such functionals~\cite{Bender2003_RMP75-121}. These methods are similar to those used in electronic systems where the form of the density functionals can be deduced {\it ab initio} from the well-known Coulomb force between the electrons~\cite{Perdew2004_LNP620-269,Bartlett2010_MPH108-3299}.
Contrary to that, at present, the nuclear density functionals are constructed phenomenologically. The form of those functionals is motivated by the symmetries of the underlying basic theories. The parameters of the model, however, are adjusted to experimental data in finite nuclei.

Within the concept of density functional theory, the full quantum-mechanical nuclear many-body problem is mapped onto a single-particle problem, assuming that the exact ground-state of the $A$-body system is determined by a Slater determinant and the corresponding single-particle density matrix generated from the products of $A$ single-particle states.
By imposing a variation principle on the energy functional with respect to this density one derives the equations of motion of the independently moving nucleons. The specific form of the phenomenological density functional leads to a certain form of the mean-field.

There are two general versions of this theory. The standard since almost fifty years ago, are nonrelativistic  functionals. The most widely known forms are the Skyrme-type functionals, based on zero-range interactions~\cite{Vautherin1972_PRC5-626} and the Gogny-type functionals of finite-range interactions~\cite{Decharge1980_PRC21-1568}. Later on, covariant density functionals were introduced. Their relativistic form is based on the simple model of Walecka~\cite{Walecka1974_APNY83-491,Serot1986_ANP16-1} and its density dependence was introduced by nonlinear meson couplings by Boguta and Bodmer~\citep{Boguta1977_NPA292-413}.

Both relativistic and nonrelativistic models have been very successful in describing bulk and structure properties of nuclei all along the beta stability line giving very similar results. However, going to nuclei close to the drip line with high isospin values there were significant differences in measuring special quantities.
A characteristic case has been the failure of the standard Skyrme functionals used at the time, to reproduce the observed kink in the radii difference of the chain of Pb isotopes, whereas relativistic functionals were very successful at reproducing it~\cite{Sharma1993_PLB317-9}. It was afterwards recognised that this qualitatively distinct result was due to the different way the two methods treat the spin-orbit interaction. In the Skyrme-Hartree-Fock (SHF) models the large spin-orbit in nuclei, which is known since the early days of the shell model~\citep{GoeppertMayer1949_PR75-1969,Haxel1949_PR75-1766}, is included phenomenologically in the form of the functional with an additional parameter that has to be adjusted to the experimental data. In contrast the covariant treatment gives rise to the very large spin-orbit coupling in a natural way. It has its origin in the fact, that the nuclear Dirac equation contains a very large attractive scalar field and a very large repulsive vector field. For the normal potential these two fields compensate to a large extent, but their effects add up in the spin-orbit term~\cite{Duerr1956_PR103-469,Miller1972_PRC5-241}.

In all the conventional nonrelativistic models the spin-orbit term is derived from a two-body spin-orbit interaction of zero range~\cite{Vautherin1972_PRC5-626,Decharge1980_PRC21-1568,Baldo2008_PLB663-390}. The corresponding Fock term leads to a strong isospin dependence of the spin-orbit splitting. This is the origin of the failure to reproduce the kink in the isotopic shifts mentioned above. In covariant models the spin-orbit splitting is a single-particle effect, derived directly from the Dirac equation. Its isospin dependence is given by the $\rho$ meson. Its strength is determined by the symmetry energy and it leads usually only to a weak isospin dependence~\cite{Duerr1956_PR103-469,Walecka1974_APNY83-491}. The use of an additional scalar isovector $\delta$ meson does not change very much this situation, because the contributions of the isovector mesons to the spin-orbit term are small as compared with the contributions of the isoscalar mesons~\cite{RocaMaza2011_PRC84-054309}.

Of course the strong isospin dependence of the spin-orbit term in conventional nonrelativistic density functionals introduced by the Fock term can be avoided, if the assumption is given up that the density functional is derived as the expectation value of an effective Hamiltonian which leads inevitably to exchange terms.
Therefore, subsequent efforts to correct for this result have led to modified Skyrme schemes where the strength of the Fock term in the functional is used as a fit parameter~\cite{Sharma1994_PRL72-1431,Reinhard1995_NPA584-467}.
In this way an extension of the Skyrme functional was proposed~\citep{Reinhard1995_NPA584-467}, reproducing also the evolution of the nuclear radius with neutron number $N$ for the isotope series of Pb and Ca. The resulting functionals were able to correct for the initial failure by changing the density dependence of the neutron spin-orbit potential.

Another example for the differences of the models in the spin-orbit part was observed in
Ref.~\citep{Lalazissis1998_PLB418-7}. It was found out, that in the framework of relativistic mean-field theory, there was a significant reduction in the spin-orbit potential in light drip line nuclei that have a large isospin value. This had an effect on the energy splittings of the same spin-orbit partners which were reduced for isotopes of Ne and Mg with the increase of neutron number. Again it was shown that a modification of the spin-orbit term in Skyrme has results similar with those of the relativistic mean-field.

There has been lately a renewed interest in experimental studies concerning the spin-orbit part of the nuclear force. In particular, two specific experiments~\citep{Burgunder2014_PRL112-042502,Mutschler2016_NatP} were recently published, where the structure of the $N = 20$ nucleus $^{34}$Si nucleus is investigated. The reason why this particular nucleus was chosen is its unique bubble structure, unveiled in earlier theoretical calculations~\citep{Grasso2009_PRC79-034318} using both relativistic and nonrelativistic models.
This bubble structure implies that there is a large central depletion in the proton density, which is due to the fact that the $2s_{1/2}$ proton state is essentially empty. This is exactly what it was shown {\it this time}
experimentally in the very recent study by Mutschler \textit{et al.} \cite{Mutschler2016_NatP}, where they used the one proton removal $(-1p)$ method, to probe the interior of the $^{34}$Si nucleus and to show that the $2s_{1/2}$ is indeed empty.

Following, therefore, the identification of $^{34}$Si as a bubble nucleus~\citep{Grasso2009_PRC79-034318} a very specific experiment by Burgunder {\textit et al.}~\cite{Burgunder2014_PRL112-042502} was conducted to attempt to set an additional constraint on the strength of the spin-orbit force. Comparing these results with earlier experiments on nuclei within the $N=20$ isotone chain, such as in Refs.~\citep{Uozumi1994_PRC50-263,Eckle1989_NPA491-205}, one was able to evaluate a reduction in the $2p_{3/2}-2p_{1/2}$ splitting. This effect has been attributed to the occurrence of a bubble in the central proton density as one advances from $^{36}$S to $^{34}$Si.
This is analogous to the case discussed in Ref. \cite{Lalazissis1998_PLB418-7} where the addition of neutrons in Ni and Sn isotopes, leads to the weakening of the spin-orbit potential and to a subsequent reduction of the size of the spin-orbit splitting of the neutron subsystem. Therefore, it has been suggested, that this kind of specific measurement could work complementary to the aforementioned theoretical studies, in order to investigate further the spin-orbit force in various mean-field models.

There has already been a study within the nonrelativistic mean-field approach~\citep{Grasso2015_PRC92-054316} where the $2p$ and $1f$ neutron spin-orbit splittings in the $N = 20$ isotones $^{40}$Ca, $^{36}$S and $^{34}$Si
have been analyzed for various Skyrme and Gogny functionals. Inspired by this work, we carried out an investigation within self-consistent covariant density functional theory describing the same nuclei as well as $^{38}$Ar. Concentrating on the first $1f_{7/2}$, $2p_{3/2}$, $2p_{1/2}$, and $1f_{5/2}$ neutron states, we calculated the SO splittings of the $ 2p $  and $ 1f $ orbitals and compared them with the corresponding nonrelativistic and experimental results.
Our goal is to examine whether the different treatments of the spin-orbit force in relativistic and nonrelativistic mean-field models gives rise to significantly different results, as it has been the case for the investigations mentioned above~\cite{Sharma1993_PLB317-9,Reinhard1995_NPA584-467,Lalazissis1998_PLB418-7}

We first neglect pairing correlations, as it has been done in the earlier nonrelativistic work of Ref.~\citep{Grasso2015_PRC92-054316} and calculate the single-particle energies in the relativistic Hartree model (RH) based on several modern nonlinear and density-dependent covariant density functionals. Afterwards we go beyond these investigations in various aspects: we study the influence of pairing correlations within the relativistic Hartree-Bogoliubov (RHB) scheme, we include tensor forces in relativistic Hartree-Fock (RHF) theory, and finally we go beyond mean-field and include particle-vibration coupling (PVC).

Our article is organized in the following way:  In Sec.~\ref{theory} we present the theoretical methods and in Sec.~\ref{extensions} we introduce specific extensions. Section~\ref{numerics} is devoted to numerical details of the calculations and in Sec.~\ref{results} we discuss the results of our investigations. Section~\ref{conclusions} contains conclusions and an outlook for future work.

\section{\label{theory}Theory}
As described in Ref.~\cite{Vretenar2005_PR409-101} in the relativistic case nucleons are treated as four-component Dirac spinors and the interaction is mediated by the exchange of virtual mesons. The minimal set of meson fields required to describe bulk and single-particle nuclear properties have the following quantum numbers and properties:
\begin{itemize}
\item[(i)] $\sigma$ meson: $J^{\pi}, T = 0^+,0,$ medium range attraction

\item[(ii)] $\omega$ meson: $J^{\pi}, T = 1^-,0,$ short range repulsion

\item[(iii)] $\rho$ meson: $J^{\pi}, T = 1^-,1,$ isospin channel.
\end{itemize}
Inspired by {\it ab initio} calculations~\cite{VanDalen2007_EPJA31-29} one has introduced in some
models in addition an isovector scalar meson, the $\delta$ meson~\cite{RocaMaza2011_PRC84-054309}:
\begin{itemize}
\item[(iv)] $\delta$ meson: $J^{\pi}, T = 0^+,1,$ isospin channel.
\end{itemize}

The model is defined by the Lagrangian density
\begin{equation}
\mathcal{L}=\mathcal{L}_{N}+\mathcal{L}_{m}+\mathcal{L}_{int}.
\label{Lagrangian}
\end{equation}%
$\mathcal{L}_{N}$ denotes the Lagrangian of the free nucleon
\begin{equation}
\mathcal{L}_{N}=\bar{\psi}\left( i\gamma ^{\mu }\partial _{\mu }-M\right)
\psi ,
\end{equation}%
where $M$ is the bare nucleon mass and $\psi $ denotes the Dirac spinor. $%
\mathcal{L}_{m}$ is the Lagrangian of the free meson fields and the
electromagnetic field
\begin{eqnarray}
\mathcal{L}_{m} &=&\frac{1}{2}\partial _{\mu }\sigma \partial ^{\mu }\sigma -%
\frac{1}{2}m_{\sigma }^{2}\sigma ^{2}+\frac{1}{2}\partial _{\mu }\vec{\delta}%
\partial ^{\mu }\vec{\delta}-\frac{1}{2}m_{\delta }^{2}\vec{\delta}^{2}
\label{L-meson}\notag\\
&&-\frac{1}{4}\Omega _{\mu \nu }\Omega ^{\mu \nu }+\frac{1}{2}m_{\omega
}^{2}\omega _{\mu }\omega ^{\mu }-\frac{1}{4}\vec{R}_{\mu \nu }\vec{R}^{\mu
\nu }+\frac{1}{2}m_{\rho }^{2}\vec{\rho}_{\mu }\vec{\rho}^{\mu }\notag\\
&&-\frac{1}{4}F_{\mu \nu }F^{\mu \nu },
\end{eqnarray}%
with the corresponding masses $m_{\sigma }$, $m_{\omega }$, $m_{\rho }$, and
$\Omega _{\mu \nu }$, $\vec{R}_{\mu \nu }$, $F_{\mu \nu }$ are the field tensors
\begin{equation}
\begin{array}{rcl}
\Omega _{\mu \nu } & = & \partial _{\mu }\omega _{\nu }-\partial _{\nu}\omega _{\mu }, \\
\vec{R}_{\mu \nu } & = & \partial _{\mu }\vec{\rho}_{\nu }-\partial _{\nu },%
\vec{\rho}_{\mu }, \\
F_{\mu \nu } & = & \partial _{\mu }{A}_{\nu }-\partial _{\nu }{A}_{\mu }.%
\end{array}%
\end{equation}%
The minimal set of interaction terms is contained in $\mathcal{L}_{int}$
\begin{eqnarray}
\mathcal{L}_{int} &=&-g_{\sigma }\bar{\psi}\psi \sigma - g_{\delta }\bar{\psi%
}\vec{\tau}\psi \vec{\delta}  \label{L-int} \\
&&-g_{\omega }\bar{\psi}\gamma ^{\mu }\psi \omega _{\mu }-g_{\rho }\bar{\psi}%
\vec{\tau}\gamma ^{\mu }\psi \cdot \vec{\rho}_{\mu }-e\bar{\psi}\gamma ^{\mu
}\psi A_{\mu } .  \notag
\end{eqnarray}%
where $e$ vanishes for neutrons. It was recognised that this linear model
was not very successful for a quantitative description of nuclei. Therefore
Boguta and Bodmer~\cite{Boguta1977_NPA292-413} introduced a density
dependence by nonlinear meson couplings replacing the quadratic term $\frac{%
1}{2}m_{\sigma }^{2}\sigma ^{2}$ by a renormalizable $\phi ^{4}$-theory
\begin{equation}
U(\sigma )=\frac{1}{2}m_{\sigma }^{2}\sigma ^{2}+\frac{1}{3}g_{2}\sigma ^{3}+%
\frac{1}{4}g_{3}\sigma ^{4}.
\end{equation}%
Later on one has also introduced nonlinear couplings in the $\omega $- and $%
\rho $-sector. As examples for such functionals we use in this investigation
the parameter set NL3~\cite{Lalazissis1997_PRC55-540}, NL3$^{*}$~\cite%
{Lalazissis2009_PLB671-36}, and FSUGold~\cite{Todd-Rutel2005_PRL95-122501}.

Through the classical variation of the Lagrangian with respect to the
different fields we find the equations of motion, the Dirac equation for the
spinors and Klein-Gordon equations for the mesons. In the static case with
time-reversal invariance have
\begin{equation}
(\bm{\alpha\cdot p}+\beta (M+S)+V)\psi _{i}=\varepsilon _{i}\psi _{i},
\label{Dirac_eq}
\end{equation}%
where the relativistic scalar and vector fields $S$ and $V$ are given by
\begin{equation}
S=g_{\sigma }\sigma +g_{\delta }\delta \text{ \ \ \ \ \ and \ \ \ \ \ }%
V=g_{\omega }\omega ^{0}+g_{\rho }\tau _{3}\rho _{3}^{0}+eA^{0}.
\label{fields}
\end{equation}

Varying the Lagrangian with respect to the meson fields we get the
Klein-Gordon type equations. Using also the Lorentz gauge for the vector
mesons they have the following form
\begin{eqnarray}
(-\Delta +m_{\sigma }^{2})\sigma &=&-g_{\sigma }\sum_{i=1}^{A}\bar{\psi}%
_{i}\psi _{i}-g_{2}\sigma ^{2}-g_{3}\sigma ^{3},  \label{Klein-Gordon1} \\
(-\Delta +m_{\delta }^{2})\delta &=&-g_{\delta }\sum_{i=1}^{A}\bar{\psi}%
_{i}\tau _{3}\psi _{i}, \\
(-\Delta +m_{\omega }^{2})\omega ^{0} &=&g_{\omega }\sum_{i=1}^{A}\psi
_{i}^{\dag }\psi _{i},  \label{Klein-Gordon2} \\
(-\Delta +m_{\rho }^{2})\rho _{3}^{0} &=&g_{\rho }\sum_{i=1}^{A}\psi
_{i}^{\dag }\tau _{3}\psi _{i},  \label{Klein-Gordon3} \\
-\Delta A^{0} &=&\frac{e}{2}\sum_{i=1}^{A}\psi _{i}^{\dag }(1-\tau _{3})\psi
_{i}.  \label{Klein-Gordon4}
\end{eqnarray}%
The sources of the fields are the various densities, such as, for instance, the
scalar density $\rho _{s}$ and the baryon density $\rho $:

\begin{equation}
\rho _{s}=\sum_{i=1}^{A}\bar{\psi}_{i}\psi _{i},\text{ \ \ \ and \ \ \ \ }%
\rho =\sum_{i=1}^{A}\psi _{i}^{\dag }\psi _{i},
\end{equation}%
and in a similar way we have the density for protons and neutrons $\rho _{n}$
and $\rho _{p}$. The summation runs always over the occupied states in the
Fermi sea (\textit{no-sea approximation}).

More modern functionals describe the density dependence not by nonlinear
meson couplings, but rather by density-dependent coupling constants: $%
g_{i}(\rho )$ (for $i=\sigma ,\delta ,\omega ,\rho $). Instead of
following the approach with nonlinear terms, an idea to use density-dependent
couplings was first proposed by Brockman and
Toki~\cite{Brockmann1992_PRL68-3408}, who derived the density dependence from
relativistic Brueckner-Hartree-Fock calculations in nuclear matter at
various densities. Modern high precision functionals use various
phenomenological forms for the density dependence, such as, for instance, the
so-called Typel-Wolter ansatz~\cite{Typel1999_NPA656-331}:
\begin{eqnarray}
g_{i}(\rho ) &=&g_{i}(\rho _{\text{sat}})f_{i}(x)~~~~~~~~~~~~~~~~~~\mathrm{%
for}~i=\sigma ,\omega,   \label{TWansatz1} \\
g_{i}(\rho ) &=&g_{i}(\rho _{\text{sat}})\exp [-a_{i}(x-1)]~~~~\mathrm{for}%
~i=\delta ,\rho,
\end{eqnarray}%
with
\begin{equation}
f_{i}(x)=a_{i}\frac{1+b_{i}(x+d_{i})^{2}}{1+c_{i}(x+d_{i})^{2}}
\label{TWansatz2}
\end{equation}%
being a function of $x=\rho /\rho _{\text{sat}}$, where $\rho _{\text{sat}}$
is the density at saturation of symmetric nuclear matter. The Typel-Wolter
ansatz is used for the density functionals DD-ME2~\cite%
{Lalazissis2005_PRC71-024312} and DD-ME$\delta$~\cite{RocaMaza2011_PRC84-054309}.

Meson exchange forces with finite meson masses are relatively complicated,
in particular for triaxially deformed nuclei or for applications of
time-dependent density functional theory for the description of excited {states}. Therefore, in analogy with the nonrelativistic Skyrme functional, one
has introduced forces with zero range, the so-called point-coupling models~\cite{Manakos1989_ZPA334-481}. These are generalizations of the Nambu-Jona Lasinio model~\cite{Nambu1961_PR122-345} including derivative terms and
density-dependent coupling constants. In this investigation we use the point-coupling functionals PC-F1~\cite{Burvenich2002_PRC65-044308} with a
polynomial density dependence and the point-coupling functional DD-PC1~\cite{Niksic2008_PRC78-034318,Daoutidis2011_PRC83-044303} with an exponential
ansatz for the density dependence.

\subsection{Isospin dependence of the spin-orbit force}
\label{Isospin dependence}
As noted in~Refs. \cite{Duerr1956_PR103-469,Miller1972_PRC5-241}, the
spin-orbit coupling arises naturally in the relativistic formalism from the
addition of the two large fields, the vector field $V$ produced mainly by
the short-range repulsion of the $\omega$ meson, and the scalar field $S$
produced mainly by the attractive $\sigma$ mesons. The isovector mesons
$\delta $ and $\rho $ contribute to the isovector dependence of the
spin-orbit splitting~\cite{RocaMaza2011_PRC84-054309}.

In the nonrelativistic expansion of the Dirac equation~\cite%
{Koepf1991_ZPA339-81} the spin-orbit term obtains the form
\begin{equation}
V_{S.O.}=\bm{W}\cdot \left( \bm{p}\times \bm{\sigma}\right),   \label{WSO}
\end{equation}%
with
\begin{equation}
\bm{W}=\frac{1}{2\tilde{M}^{2}}\bm{\nabla}(V-S)
\end{equation}%
and the effective mass
\begin{equation}
\tilde{M}=M-\frac{1}{2}(V-S).
\end{equation}%
In the spherical case we have%
\begin{equation}
V_{S.O.}=\frac{1}{4\tilde{M}^{2}}\frac{1}{r}\frac{d(V-S)}{dr}\bm{\ell}\cdot\bm{s}.
\end{equation}

To have a rough estimate for the isospin dependence we make the
following approximations: (i) we neglect nonlinear meson couplings as well
as the density dependence of the coupling constants, (ii) we neglect the
difference between scalar and vector density, and (iii) we solve the
Klein-Gordon equations in the local density approximation, i.e., we neglect the
Laplacians.

We thus obtain for the meson-coupling models with $C_{i}=g_{i}^{2}/m_{i}^{2}$,%
\begin{equation}
V-S=(C_{\omega }+C_{\sigma })(\rho _{p}+\rho _{n})+\tau _{3}(C_{\rho
}+C_{\delta })(\rho _{p}-\rho _{n}),
\end{equation}%
where for the meson-coupling models $C_{i}=g_{i}^{2}/m_{i}^{2}$ ($i=\sigma
,\omega ,\delta ,\rho $) and for the point-coupling models $C_{i}=\alpha
_{S},\alpha _{V},\alpha _{TS},\alpha _{TV}$. This leads to
\begin{equation}
\bm{W}_{\tau }=W_{1}\bm{\nabla}\rho _{\tau }+W_{2}\bm{\nabla}\rho _{\tau
^{\prime }\neq \tau },
\end{equation}
with $W_{1}$ very close to $W_{2}$:
\begin{equation}
\frac{W_{1}}{W_{2}}\approx 1+2\frac{C_{\rho }+C_{\delta }}{C_{\omega
}+C_{\sigma }}.
\label{W1/W2_approx}
\end{equation}%
Of course, there is also a small isospin-dependence in the effective mass
$\tilde{M}$ and, because of the density dependence, these parameters depend
on $r$. \ However, in the relevant region, for all the models, the isovector
coupling constants $C_{\rho }+C_{\delta }$ reach only 10\%-20\% of the
isoscalar values.

In principle the fit to experimental data in finite nuclei only allows us
to determine $C_{\rho }-C_{\delta }$ and not $C_{\rho }$ and $C_{\delta }$
independently~\cite{RocaMaza2011_PRC84-054309}. Therefore the $\delta$ meson is neglected in
most of the successful parameter sets ($C_{\delta }=0$). In principle $%
C_{\rho }+C_{\delta }$ could have a large value, as it happens in the
isoscalar case with the extremely large scalar and vector potentials $S$ and
$V$, which cancel in the normal mean field, but add up in the spin-orbit
term. There are, however, strong indications from {\it ab initio} calculations
that this is not the case. In fact, in the parameter set DD-ME$\delta$~\cite{RocaMaza2011_PRC84-054309} the coupling $g_{\delta }(\rho )$ was adjusted to the splitting
of the effective Dirac mass between protons and neutrons, as has been
calculated in relativistic Brueckner-Hartree-Fock calculations in nuclear
matter by the Tuebingen group~\cite{VanDalen2007_EPJA31-29}.

In the nonrelativistic density functionals of Skyrme and Gogny type the
spin-orbit term is derived from a zero-range two-body spin-orbit interaction
of the form
\begin{equation}
V_{12}^{(SO)}(\bm{r}_{12}) = iW_{0}(\bm{\sigma_1}+\bm{\sigma_2})\cdot (%
\bm{\hat{k}}^{\dagger }\times \delta (\bm{r}_{12})\bm{\hat{k}}),
\label{Skyrme_force}
\end{equation}%
with $\bm{r}_{12}=\bm{r}_{1}-\bm{r}_{2}$, and $\bm{\hat{k}}%
=-(i/2)(\nabla _{1}-\nabla _{2})$. The parameter $W_{0}$, together with
the remaining parameters, is determined phenomenologically through a fit to
finite nuclei. Since these are Hartree-Fock calculations, the exchange term
leads to a very specific isospin dependence and the spin-orbit term has the
form of Eq. (\ref{WSO}) with
\begin{equation}
\bm{W}_{\tau }(\bm{r})=W_{1}\bm{\nabla}\rho _{\tau }+W_{2}\bm{\nabla}%
\rho _{\tau ^{\prime }\neq \tau }.  \label{WSOSkyrme}
\end{equation}%
here the parameters $W_{1}$ and $W_{2}$ are constants and because of the
exchange term one finds
\begin{equation}
\frac{W_{1}}{W_{2}}=2.
\end{equation}%
As we see in this
standard formulation of nonrelativistic forces, there is no explicit
isospin or density dependence in the spin-orbit term, but the exchange part
of the force introduces a strong
isospin dependence because of the isospin exchange operator $\hat{P%
}^{\tau }=\frac{1}{2}(1+\hat{\tau}_{1}\cdot \hat{\tau}_{2})$ .
It has been found that this particular property of the SO term leads to
considerable problems in reproducing the isotope shifts in nuclear charge
radii in the Pb region (see Refs.~\cite{Sharma1993_PLB317-9,Reinhard1995_NPA584-467}), which is not the case for the relativistic models.
Of course, density functional theory does not necessarily have to start with
a Hamiltonian treated in the Hartree-Fock approximation. In principle one can
also use general density functionals, where the exchange contribution
contains a free parameter $x_{\text{W}}$. In this case the density
functional, i.e., the expectation value of the energy, is determined in
the Hartree-approximation from a slightly modified spin-orbit term~\citep{Sharma1995_PRL74-3744,Reinhard1995_NPA584-467}
\begin{equation}
V_{SO}=iW_{0}\frac{1}{2}(1+x_{\text{w}}\hat{P}^{\tau })(\bm{\sigma}_{1}+%
\bm{\sigma}_{2})\bm{\hat{k}^{\dagger }}\times \delta (\bm{r}_{12})%
\bm{\hat{k}}.  \label{mSkA}
\end{equation}%
When the single-particle field is derived from this functional we end up
with a spin-orbit potential of the form (\ref{WSOSkyrme}) with $%
W_{1}\,=\,W_{0}(1+x_{w})/2$, $W_{2}\,=\,W_{0}/2$. Using the modified Skyrme
ansatz there is the ability to allow for change in the isospin-dependence of
Skyrme forces through the parameter $x_{w}$~\citep{Sharma1995_PRL74-3744,Reinhard1995_NPA584-467}.
With this kind of modification one was able to reproduce the kink isotopic shifts of Pb nuclei.

\subsection{\label{pairing}Pairing correlations}

The theory we have presented above remains in the relativistic mean-field
level and since we neglect any exchange terms we have a relativistic Hartree
approximation to describe the long-range particle-hole correlations in a
nucleus. However, in open-shell nuclei we know that particle-particle
correlations are important and one should have to take them into account
explicitly. In the nonrelativistic functionals this is done in the
Hartree-Fock-Bogoliubov (HFB) theory~\citep{Bulgac1980_nucl-th9907088,Dobaczewski1996_PRC53-2809}
that provides a unified picture for the mean-field and pairing correlations. The
relativistic version of the transformation is a hybrid where the long-range
interaction is given by the Lorenz-covariant Lagrangians, we have given
above, and the short-range interaction is produced by effective
nonrelativistic forces. Pairing correlations can be easily included in the
framework of density functional theory, by using a generalized Slater
determinant $|\Phi \rangle $ of the Hartree-Bogoliubov type. The ground
state of a nucleus $|\Phi \rangle $ is represented as the vacuum with
respect to independent quasiparticle operators
\begin{equation}
\alpha _{k}^{+}=\sum\limits_{l}U_{lk}c_{l}^{+}+V_{lk}c_{l}^{{}},
\end{equation}%
where $U_{lk}$, $V_{lk}$ are the Hartree-Bogoliubov coefficients. They
determine the Hermitian single-particle density matrix%
\begin{equation}
\hat{\rho} = V^{*}V^{T},\;  \label{rho0}
\end{equation}%
and the antisymmetric pairing tensor
\begin{equation}
\hat{\kappa} = V^{*}U^{T}.  \label{kappa0}
\end{equation}%
The energy functional depends not only on the density matrix $\hat{\rho}$
and the meson fields $\phi _{m}$, but also on the pairing tensor.
\begin{equation}
E[\hat{\rho},\hat{\kappa},\phi_m ]=E_{RMF}[\hat{\rho},\phi_m ]+E_{pair}[\hat{%
\kappa}],  \label{ERHB}
\end{equation}%
where $E_{RMF}[\hat{\rho},\phi ]$ is the $RMF$-functional. The pairing
energy $E_{pair}[\hat{\kappa}]$ is given by%
\begin{equation}
E_{pair}[\hat{\kappa}]=\frac{1}{4}\text{Tr}\left[ \hat{\kappa}^{\ast }V^{pp}%
\hat{\kappa}\right] .
\end{equation}%
$V^{pp}$ is a general two-body pairing interaction.

To get a static solution for ground states of open-shell nuclei in this
framework, we have to solve the Hartree-Bogoliubov equations%
\begin{equation}
\left(
\begin{array}{cc}
\hat{h}-m-\lambda & \hat{\Delta} \\
-\hat{\Delta}^{\ast } & -\hat{h}+m+\lambda%
\end{array}%
\right) \left(
\begin{array}{c}
U_{k}(\bm{r}) \\
V_{k}(\bm{r})%
\end{array}%
\right) =E_{k}\left(
\begin{array}{c}
U_{k}(\bm{r}) \\
V_{k}(\bm{r})%
\end{array}%
\right) \;.  \label{eqhb}
\end{equation}

This system of equations contains two average potentials: the
self-consistent mean field $\hat{h}$, which encloses all the long range
particle-hole (\textit{ph}) correlations, and the pairing field $\hat{\Delta}
$, which includes the particle-particle (\textit{pp}) correlations. The
single-particle potential $\hat{h}$ results from the variation of the
energy functional with respect to the Hermitian density matrix $\hat{\rho}$,
\begin{equation}
\hat{h}=\frac{\delta E}{\delta\hat{\rho}},
\end{equation}%
and the pairing field is obtained from the variation of the energy
functional with respect to the pairing tensor
\begin{equation}
\hat{\Delta}~=~\frac{\delta E}{\delta \hat{\kappa}}.
\end{equation}

The chemical potential $\lambda$ is determined by the particle number
subsidiary condition in order that the expectation value of the particle
number operator in the ground state equals the number of nucleons. The
column vectors denote the quasiparticle wave functions, and $E_k$ are the
quasiparticle energies. The dimension of the RHB matrix equation is two
times the dimension of the corresponding Dirac equation. For each
eigenvector $(U_k ,V_k )$ with positive quasiparticle energy $E_k > 0$,
there exists an eigenvector $(V_k^*,U_k^*)$ with quasiparticle energy $-E_k$.
Since the baryon quasiparticle operators satisfy fermion commutation
relations, the levels $E_k$ and $-E_k$ cannot be occupied simultaneously.
For the solution that corresponds to a ground state of a nucleus with even
particle number, one usually chooses the eigenvectors with positive
eigenvalues $E_k$.

The eigensolutions of Eq. (\ref{eqhb}) form a set of orthogonal
(normalized) single quasiparticle states. The corresponding eigenvalues are
the single quasiparticle energies. The self-consistent iteration procedure
is performed in the basis of quasiparticle states. The resulting RHB-function
is analyzed in the canonical basis~\cite{Ring1980}, where it has the form of
a BCS-function. In this basis the density matrix
$R_{kk^{\prime }}=\bigl< V_k(\bm{r})\big\vert V_{k^{\prime }}(\bm{r})\bigl>$
is diagonal and its eigenvalues are the BCS-occupation probabilities
\begin{equation}
v_\mu^2 = \frac{1}{2}\left[1 - \frac{\varepsilon_\mu - \lambda}{\sqrt{(\varepsilon_\mu -
\lambda)^2 + \Delta_\mu^2}}\right].
\label{BCS}
\end{equation}
Here the $\varepsilon_\mu=\langle\mu|\hat{h}|\mu\rangle$ are the single-particle
energies in the canonical basis and $\Delta_\mu=\langle\mu|\hat{\Delta}|\bar{\mu}\rangle$
are the corresponding gap-parameters.

If the pairing field $\hat \Delta$ is diagonal and constant, HFB reduces to
the BCS-approximation. The lower and upper components $U_k(\bm{r})$ and $%
V_k(\bm{r})$ are equivalent, with the BCS-occupation amplitudes $u_k$
and $v_k$ as proportionality constants. In that case we use the odd-even
mass difference to obtain the value of experimental gap parameter
\begin{equation}  \label{OES3pt-standard}
\Delta = \frac{(-1)^{N+1}}{2} \left[E(N+2) - 2E(N+1) + E(N))\right]
\end{equation}
and the occupation probabilities are given by the BCS formula (\ref{BCS}).

The problem with this simplified method is the ultraviolet divergence of the
pairing field for high momenta. This means that it is necessary to have a
fixed pairing window or an energy cut-off, which adds an extra parameter in
the model that cannot be fixed experimentally.

This can be avoided for finite range effective pairing forces. One way
suggested in Ref.~\cite{GonzalesLlarena1996_PLB379-13} is using a nonrelativistic
pairing interaction based on the pairing part of the well known and very
successful Gogny force~\citep{Berger1984_NPA428-23},
\begin{eqnarray}
V^{pp}(1,2)&=& \sum_{i=1,2}e^{-((\bm{r}_{1}-\bm{r}_{2})/{\mu _{i}}%
^{2}}  \notag \\
&&\times(W_{i}+B_{i}P^{\sigma }-H_{i}P^{\tau }-M_{i}P^{\sigma }P^{\tau }),
\end{eqnarray}
with the set D1S~\cite{Berger1984_NPA428-23} for the parameters $\mu _{i}$, $W_{i}$,
$B_{i}$, $H_{i}$, and $M_{i}$ $(i=1,2)$. This force has been very carefully
adjusted to the pairing properties of finite nuclei all over the periodic
table. In particular, the basic advantage of the Gogny force is the finite
range, which automatically guarantees a proper cut-off in momentum space.

This method has been very successful but it requires great computational
effort. So an alternative was developed in Ref.~\cite{TIAN-Y2009_PLB676-44} by Tian {\textit et al.}
 (TMR), that has been formulated as a separable force in momentum
space. Therefore it can be determined by two parameters adjusted to
reproduce the pairing gap of the Gogny force in symmetric nuclear matter. In
the $^1S_0$ channel the gap equation reads
\begin{equation}
\Delta(k) = - \int_0^{\infty} \frac{k^{\prime 2}dk^{\prime }}{2\pi^2}
\langle{}k\vert{}V^{^1S_0}\vert{}k^{\prime }\rangle \frac{\Delta(k^{\prime })%
}{2E(k^{\prime })},
\end{equation}%
and the pairing force separable in momentum space is
\begin{equation}
\langle{}k\vert{}V^{^1S_0}\vert{}k^{\prime }\rangle = - Gp(k)p(k^{\prime }).
\end{equation}
The two parameters determining the force are, the pairing strength $G$ and $%
\alpha$ that goes in the Gaussian ansatz $p(k) = e^{-\alpha^2k^2}$. Their
value has been adjusted to $G = 728$ MeV fm$^3$ and $\alpha = 0.644$ fm in
order to reproduce the density dependence of the gap at the Fermi surface,
calculated with the D1S parametrization of the Gogny
force~\cite{Berger1991_CPC63-365}.

\section{\label{extensions}Specific extensions}

\subsection{\label{tensor-force}Tensor forces}

It is generally acknowledged that the tensor part of the nuclear force plays an essential role in the description of the several nuclear properties. The standard formulation of covariant density functionals is based on the Relativistic Hartree approximation, i.e., exchange terms are not taken into account explicitly. This is in most cases a good approximation, because the coupling constants in the various spin-isospin channels are adjusted to experimental data. For zero-range forces the Fierz theorem shows, that exchange terms can be expanded over direct terms with new effective coupling constants being linear combinations of the old coupling constants in the different spin-isospin channels. For meson-exchange forces with heavy meson masses, such as the $\sigma$-, $\omega$-, $\delta$- and $\rho$ mesons, the corresponding ranges are short and therefore this is still a reasonable approximation.

Following this arguments, in conventional covariant density functional theory the contributions of the exchange terms are taken into account effectively through the adjustment of the parameters to experimental data.
As already mentioned, this model has been extremely successful in describing a vast range of
nuclear bulk properties, such as binding energies, radii, deformation parameters, giant resonances,
etc.~\cite{Meng2016_IRNP10}.

Of course, the pion mass is small and, therefore, its exchange term should be taken into
account explicitly. There have been also recent studies, which found that the inclusion of
a tensor force has an effect on very specific single-particle observables. It has been shown,
for instance, in Ref.~\cite{Otsuka2005_PRL95-232502}, that
tensor forces are responsible for the shift of effective single-particle
levels in shell-model calculations for exotic nuclei. Furthermore, the
spin-orbit alignment is crucial to the strongly repulsive or attractive
character of the tensor force between proton or neutrons. So, in our case,
where we want to study the spin-orbit coupling, the effect of tensor forces
may prove to have some quantitative importance, as this has also been
investigated in the nonrelativistic study in Ref.~\cite{Grasso2015_PRC92-054316}.

In the relativistic scheme tensor terms usually show up, if one takes into
account exchange terms. Relativistic Hartree-Fock (RHF) theory including tensor
terms has a long history~\cite{Brockmann1978_PRC18-1510,Bouyssy1987_PRC36-380,Bernardos1993_PRC48-2665},
but such calculations require a considerable computational effort. Therefore,
for a long time computer power was too limited to determine
in a consistent and successful way the parameters of a relativistic density
functional containing tensor terms. In the mean time two groups have overcome
these problems. Long {\textit et al.} developed a spherical
RHF-code~\cite{PhD_Long2005,Long2006_PLB640-150,Long2007_PRC76-034314}
in $r$-space containing all the exchange terms for the $\sigma$-,  $\omega$-,
and $\rho$ mesons and for the $\pi$ meson with density-dependent coupling constants.
By adjusting the
corresponding parameters to the usual data of binding energies and radii
in finite nuclei they determined several successful parameter sets
for the RHF description of nuclei all over the periodic table.
Serra {\textit et al.}~\cite{PhD_Serra2001,Lalazissis2009_PRC80-041301} developed an RHF-code
in oscillator space taking into account only the exchange term of the $\pi$ meson
because the other mesons $\sigma$,  $\omega$, and $\rho$ are relatively heavy and
the corresponding force is of short range. Therefore, as discussed before, the
exchange terms of these mesons can be represented in the static case to a good
approximation by direct terms with effective coupling constants.

In this work we follow this method to take into account tensor terms in the
relativistic scheme. Basically two terms are added in the Lagrangian of the system,
the first is the term of the free pion field included in $\mathcal{L}_m$ as given in Eq. (\ref{L-meson}),
\begin{equation}
\mathcal{L}_{\pi} = \frac{1}{2}\left(\partial_{\mu}\vec{\pi}\partial^{\mu}%
\vec{\pi} - m^2_{\pi}\right)\vec{\pi}^2,
\end{equation}
where the mass of the pion is set to its experimental value $m_{\pi}= 138$
MeV. The second term is the pseudo-vector Yukawa type of force included in $%
\mathcal{L}_{int}$ as given in Eq. (\ref{L-int}),
\begin{equation}
\mathcal{L}_{pv} = - \frac{f_{\pi}}{m_{\pi}}\bar{\psi}\gamma_{5}\gamma_{%
\mu}\partial^{\mu}\vec{\pi}\vec{\tau}\psi,
\end{equation}
$f^2_{\pi}=\lambda f^{2 free}_{\pi}$ is the strength of the one-pion-exchange
interaction in this model and $f^{free}_{\pi}$ is the experimental value of
pion-nucleon coupling in free space.
A factor $\sqrt{\lambda}$ is used as a multiplier to vary the coupling constant
of the pion from zero ($\lambda$=0) to its free value $f^{free}_{\pi}$ ($\lambda$=1).
This comprises now a Relativistic Hartree-Fock model and its parameters have been
readjusted for different values of $\lambda$. This has been done following the same
procedure that was used to adjust the parameters of NL3~\cite{Lalazissis1997_PRC55-540}.

Concentrating in this fit only to binding energies and radii of finite nuclei,
it was shown that the optimal fit was achieved for $\lambda=0$,
i.e., for vanishing pion-nucleon interaction. However, a parameter set NL3RHF0.5 with half the
strength of the free pion ($\lambda$=0.5) describes in addition to the other
data the evolution of single-particle structure in the tin isotopes measured by
the Argonne group~\cite{Schiffer2004_PRL92-162501} in ($\alpha$,t)
transfer reactions .

\subsection{Particle-Vibrational Coupling}
\label{PVC}
So far we discussed only mean-field methods to describe single-particle
energies. In this description of the nuclear many-body system the nucleons
move independently. In the next step we go beyond the mean-field description
and include correlations by the method of particle-vibration coupling (PVC).
This is important for our investigation of single-particle excitations,
since the coupling of the single-particle motion to the low-lying phonons
leads to a fragmentation of the single-particle spectrum, a feature most
prominent in spherical nuclei~\cite{Litvinova2011_PRC84-014305}. Even though
conventional DFT reproduce fairly well the gross structure of the SO
splitting, the inclusion of particle-vibration coupling produces a denser
spectrum near the Fermi surface which is in better agreement with
experimental observations.

In fact, it is well known from Landau-Migdal theory~\cite{Landau1959_JETP8-70,Migdal1967} that particles in the many-body system can interact with low-lying surface phonons and form
Landau quasiparticles surrounded by a cloud of excitons. Such phenomena lead
to a fragmentation of the single-particle energies. In DFT such effects can be taken into account in the
framework of time-dependent density functional theory (TDDFT)~\cite{Marques2006}.
In contrast to static DFT, which depends only on the exact
static density $\rho _{0}(\mathbf{r})$, its basis is the exact
time-dependent density $\rho (\mathbf{r,}t)$, which depends on four
variables.
In static Kohn-Sham theory the static
density $\rho _{0}(\mathbf{r})$ is mapped onto a static single-particle
potential, the Kohn Sham potential or the static self-energy  $\Sigma _{%
\text{KS}}$, which is easy to diagonalize and whose local single-particle
density is identical to the exact local ground-state density $\rho _{0}(%
\mathbf{r})$.
In full analogy to the static DFT, in the time-dependent case there exists a time-dependent
single-particle field, the time-dependent self-energy $\Sigma (\mathbf{r},t)$
with a time-dependent density identical to the exact local single-particle
density $\rho (\mathbf{r,}t)$ of the time-dependent many-body problem. This
is the Runge-Gross theorem~\cite{Runge1984_PRL52-997}. The problem is, that
we know very little about this time-dependent self-energy. It is very
complicated because it contains all the memory effects of the system.

In the case, where the time-dependent motion is
of a small-amplitude character,
one can apply linear-response theory and determine the time-dependent
self-energy in a perturbative approach.
In Fourier space one ends up with a self-energy depending on the energy $%
\omega $.  Green's function techniques and diagrammatic expansions are used to
provide a model for the energy-dependent self-energy $\Sigma (\mathbf{r}%
,\omega )$.

In a first step one starts with the ground state of the even system
determined by static DFT and allows for small amplitude vibrations around
this static solution. In the \textit{adiabatic approximation} one assumes
that, at each time, the self-energy is identical to the static self-energy
calculated with the density $\rho (\mathbf{r,}t)$. This leads to
time-dependent mean-field theory and in the limit of small amplitudes to the
well-known random-phase approximation (RPA), in the relativistic case to
relativistic RPA (RRPA), and in the case of pairing to quasiparticle RPA
(QRPA). In this way one calculates collective excitations such as, for instance,
the surface phonons, which are linear superpositions of $ph$ excitations, by
the diagonalization of the RPA-matrix. The interaction between these
$ph$-pairs is given as the second derivative of the energy density functional
with respect to the density%
\begin{equation}
V(\mathbf{r}_{1},\mathbf{r}_{2})=\frac{\delta ^{2}E[\rho ]}{\delta \rho (%
\mathbf{r}_{1})\delta \rho (\mathbf{r}_{2})}.  \label{eq:Vstatic}
\end{equation}
One obtains harmonic vibrations $|\mu \rangle $ with the eigenfrequencies $%
\Omega _{\mu }$ and the transition densities $\delta \rho _{12}^{\mu
}=\langle \mu |a_{2}^{\dag }a_{1}^{{}}|0\rangle .$

In the next step one goes back to the description of single-particle motion
in the presence of the collective vibrations. Starting from the static mean
field in the self-energy, one adds terms which describe the coupling of
single-particle motion to the vibrations. These terms are energy-dependent. The
coupling is provided by the vertices of the form
\begin{equation}
\gamma _{12}^{\mu }=\sum_{34}\langle 14|V|23\rangle \delta \rho _{34}^{\mu },
\end{equation}%
where $\langle 14|V|23\rangle $ are the matrix elements of the interaction (%
\ref{eq:Vstatic}) and $\delta \rho _{34}^{\mu }$ are the transition densities of
the corresponding phonons.

Finally, the energy-dependent part of the self-energy $\Sigma (\omega )$ is
found in second order of the particle-vibration coupling:
\begin{equation}
\Sigma _{12}^{(e)}(\omega )=\sum_{k\mu }\left( \frac{\gamma _{1k}^{\mu
}\gamma _{2k}^{\mu \ast }}{\omega -\varepsilon _{k}-\Omega ^{\mu }+i
\eta } + \frac{\gamma _{k1}^{\mu }\gamma _{k2}^{\mu \ast }}{\omega
- \varepsilon _{k}+\Omega ^{\mu }-i \eta }\right),
\label{sigmae}
\end{equation}%
where a virtual phonon with the
frequency $\Omega _{\mu }$ is emitted, moving the particle from level 1 to level
$k$.
More details can be found in
Refs.~\cite{Litvinova2006_PRC73-044328,Litvinova2007_PRC75-064308} as well as an extension of
the approach to superfluid systems in Refs. \cite{Litvinova2008_PRC78-014312,Litvinova2012_PRC85-021303}.

Combining this energy-dependent part of the self-energy with the static part
we obtain the full self-energy $\Sigma $. It contains all the forces that
act on a single nucleon. It is non-local in space and time coordinates which gives rise to an energy dependence of its Fourier transform:
\begin{equation}
\Sigma (\bm{r},\bm{r}^{\prime };\omega )=\tilde{\Sigma}(\bm{r})\delta (\bm{r}%
-\bm{r}^{\prime })+\Sigma ^{(e)}(\bm{r},\bm{r}^{\prime };\omega ),
\label{full_self_energy}
\end{equation}
where the static part $\tilde{\Sigma}$ of the self-energy, i.e. the
Dirac Hamiltonian of the ground state reads:
\begin{equation}
\hat{h}=\bm{\alpha p}+\beta (m+S)+V = \bm{\alpha p}+\beta m + \tilde{\Sigma}.
\end{equation}
This leads to the Dyson equation describing the motion of the quasiparticles
in the presence of the vibrating mean field. It can be written in terms of
Green's function as
\begin{equation}
(\varepsilon -\hat{h}-\Sigma ^{e}(\varepsilon ))G(\varepsilon )=\beta,
\end{equation}%
and in the Dirac basis, which diagonalizes the energy-independent part of
the Dirac equation, it is rewritten as follows:
\begin{equation}
\sum_{l}\{(\varepsilon -\varepsilon _{k})\delta _{{kl}}-\Sigma _{kl}^{e}(\varepsilon
)\}G_{lk^{\prime }}(\varepsilon )=\delta _{kk^{\prime }}.
\end{equation}%
In the diagonal approximation it has the form
\begin{equation}
(\varepsilon -\varepsilon _{k}-\Sigma _{k}^{e}(\varepsilon ))G_{k}(\varepsilon )=1.
\label{Dyson}
\end{equation}

For each quantum number $k$, there exist several solutions $\varepsilon
_{k}^{(\lambda )}$, which are characterized by the index $\lambda $. So the inclusion
of a coupling between the single-particle states and the vibrations leads to the fragmentation of
each single-particle state $k$.
Taking into account the pole structure of the self-energy (\ref{sigmae}), we get as an outcome an effective spectroscopic
factor $S_{k}^{(\lambda )}$ which
determines the occupation probability for each fragment $\lambda$ of the state $k$.

\section{\label{numerics}Numerical Details}

Throughout this work, the Dirac equation (\ref{Dirac_eq}) and the
Klein-Gordon equations (\ref{Klein-Gordon1})-(\ref{Klein-Gordon4}) are
solved by an expansion of the large and small components of the Dirac
spinors and of the meson fields in a spherical oscillator basis (see Ref.~\cite{Gambhir1990_APNY198-132}) with the frequency $\hslash \omega=41A^{-1/3}$. Since these eigenfunctions form an infinite set it is
necessary to truncate this basis to $N_{F}=N_{B}=20$ major oscillator shells
for the fermion and the meson fields respectively. As explained in the
following, in order to study states that belong to the continuum, we have
been changing the number $N_{F}$ to vary from $N_{F}=14-20$ shells.

In the case of tensor forces (see Sec. \ref{tensor-force}) the
Dirac-Hartree-Fock equations are solved in the same spherical oscillator
basis with $N_{F}=20$ major oscillator shells and, as discussed in Refs.~\cite{PhD_Serra2001,Lalazissis2009_PRC80-041301}, the matrix elements of the exchange term are evaluated in this basis.

For the calculations with particle-vibration coupling a seniority zero
pairing force was used. In this case the pairing potential is a multiple of
the unity matrix and the RHB-equations are identical to the RMF+BCS
equations. The strength of the coupling constant of the pairing force is
adjusted in such a way, that the resulting gap parameter is $\Delta =2$ MeV which
is close to its empirical value.
Further details of the particle-vibration coupling are given in
Refs.~\cite{Litvinova2006_PRC73-044328,Litvinova2008_PRC78-014312,Litvinova2012_PRC85-021303}.
Non-spin-flip phonons $\mu$ with natural parities, angular momenta $L_{\mu}\leq 6$, and frequencies
$\Omega^{\mu} \leq$ 20 MeV have been included into the self-energy (\ref{sigmae}) if their reduced transition probabilities exceed 5\% of the maximal ones for each $L_{\mu}$. This has been established as a standard truncation scheme for the
relativistic PVC calculations.

\section{\label{results} Results}

\begin{figure}
\includegraphics[width=1.0\columnwidth]{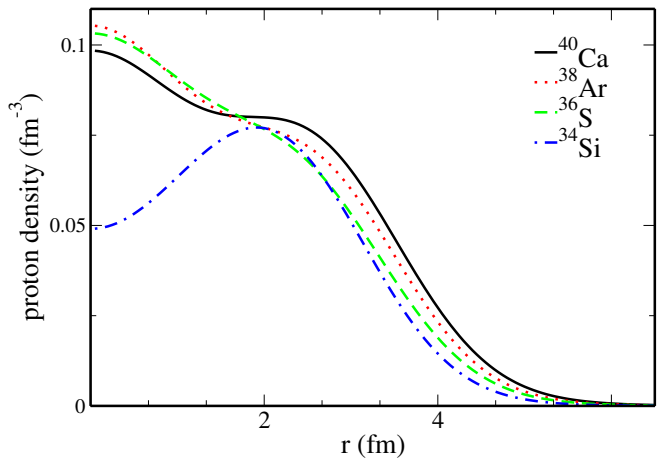}
\caption{(color online) Proton densities of the nuclei $^{40}$Ca, $^{38}$Ar, $^{36}$S and $^{34}$Si for the functional DD-ME2.}
\label{fig1}
\end{figure}

As mentioned in the introduction we concentrate our study to the series of $N = 20$ isotones. We start with the nucleus $^{40}$Ca with $Z = 20$ protons, where the last four protons fill the $1d_{3/2}$ orbit. By removing two protons we go to $^{38}$Ar and by removing two more we reach $^{36}$S which has its last two protons in the $2s_{1/2}$ orbit. The density distribution of this state is peaked in the center of the nucleus. and the removal of the two protons, as we go to $^{34}$Si, leads to an occupation probability close to zero. Therefore we have a central depletion in the proton density and the formation of a dimple around the center of the nuclear charge density. This is shown in Fig. \ref{fig1}, where we plot the proton densities with respect to the nuclear radius. For the first three nuclei in this chain we can see clearly a peak of the proton density at the center of the nucleus whereas for $^{34}$Si there is a dimple, (see also Ref.~\cite{Grasso2009_PRC79-034318}).

Experimental evidence of the existence of this bubble
structure has been given very recently by Mutschler \textit{et al.} in Ref.~\cite{Mutschler2016_NatP}, where the one-proton removal reaction $^{34}$Si$(-1p)$ $^{33}$Al has been studied.
Even though the occupancy of a single-particle orbit is not a direct
observable, its value can be calculated by using experimental data, as
it is explained  in the methods  section of that reference.
Therefore, an occupancy of 0.17(3) has been deduced for the $2s_{1/2}$ proton
state in $^{34}$Si, which is only 10\% of the 1.7(4)
occupancy of the same state in $^{36}$S, resulting in an
occupancy change of $\Delta(2s1/2) = 1.53$.

This result came in addition to the findings of the earlier
experiment by Burgunder \textit{et. al.}~\cite{Burgunder2014_PRL112-042502}, where the energies and spectroscopic factors of the first $1f_{7/2}$, $2p_{3/2}$, $2p_{1/2}$, and $1f_{5/2}$ neutron states in the  nucleus $^{35}$Si were measured through a ($d,p$) transfer reaction. Together with the results of Refs.~\citep{Uozumi1994_PRC50-263,Eckle1989_NPA491-205}, it was discovered that the
$2p = 2p_{1/2} - 2p_{3/2}$ spin-orbit splitting was considerably reduced as one goes from $^{36}$S to $^{34}$Si.

An important aspect of the spin-orbit force is its density and isospin dependence. It is clearly stated in Refs.~\citep{Burgunder2014_PRL112-042502,Mutschler2016_NatP}
that the results of these two experiments are ideal for a further theoretical investigation of the SO force deduced from the various nuclear density functionals. In particular, the extreme neutron-to-proton density asymmetry in the case of $^{34}$Si and the subsequent large and abrupt reduction in the size of the \textit{p} spitting, can provide a better constraint of the SO force, since these results isolate the contributions coming mostly from its density and its isospin dependence.

As discussed in the Sec. \ref{Isospin dependence}, the way the spin-orbit force is included in relativistic density functional theory is substantially different from the nonrelativistic case.
The ratio $W_1/W_2$ plays an important role. In the first case this ratio is density-dependent and has a value close to one, whereas for the latter case it has a fixed value equal to two. As we have already noted, this is the main reason why we get such different results in the calculations of the spin-orbit splitting.

Nonrelativistic investigations have been carried out in Ref.~\cite{Grasso2015_PRC92-054316} for the Skyrme SLy5~\cite{Chabanat1998_NPA635-231} and Gogny D1S~\cite{Berger1991_CPC63-365} functionals and certain tensor extensions of those functionals. The neutron {\it f} and {\it p} splittings for the nuclei $^{40}$Ca, $^{36}$S, and $^{34}$Si, were studied for the pure mean-field Hartree-Fock level. The corresponding results are shown in Table~\ref{tab1}.
%
\begin{table}
\renewcommand\arraystretch{1.5}
\begin{tabular}{@{}cccccccccc@{}}
\hline
\toprule
\hline
        & \multicolumn{2}{c}{$^{40}$Ca}  && \multicolumn{2}{c}{$^{36}$S}   && \multicolumn{2}{c}{$^{34}$Si}  \\
\hline\midrule\hline
Splitting & \multicolumn{1}{c} {\it f}  & \multicolumn{1}{c}  {\it p}  && \multicolumn{1}{c} {\it f}  & \multicolumn{1}{c} {\it p}  && \multicolumn{1}{c} {\it f}  & \multicolumn{1}{c} {\it p}  & \\
\hline
SLy5     & 8.39  & 2.19 && 7.88  & 2.01  && 5.86  & 1.21 &  \\
D1S    & 8.66  & 2.16  && 7.98  & 1.88 && 6.37 & 1.07  & \\
\hline
\bottomrule
\hline
\end{tabular}
\begin{tabular}{@{}ccccccc@{}}
\hline
\toprule
\hline
 &\multicolumn{2}{c}{$^{40}$Ca $\rightarrow$ $^{36}$S} & \ \ \ \  & \multicolumn{2}{c}{$^{36}$S $\rightarrow$ $^{34}$Si}&
 \\
\hline
\midrule
\hline
Splitting  & $f$ & $p$   && $f$ & $p$ & \\
\hline
SLy5     & 6\% & 8\%  &&  26\% & 40\% & \\
D1S      & 8\% & 13\% &&  20\% & 43\%& \\
\hline
\bottomrule
\hline
\end{tabular}
\caption{Sizes and relative reductions of neutron $p$ and $f$ splittings for the nonrelativistic case as, given in Ref.~\citep{Grasso2015_PRC92-054316}.}
\label{tab1}
\end{table}

Following the above experimental and theoretical studies we calculate the energies of the same neutron states and also the occupation probabilities of the $2s_{1/2}$ proton state in $^{36}$S and in $^{34}$Si for several covariant density functionals.

\begin{figure}[h!]
\includegraphics[width=1.0\columnwidth]{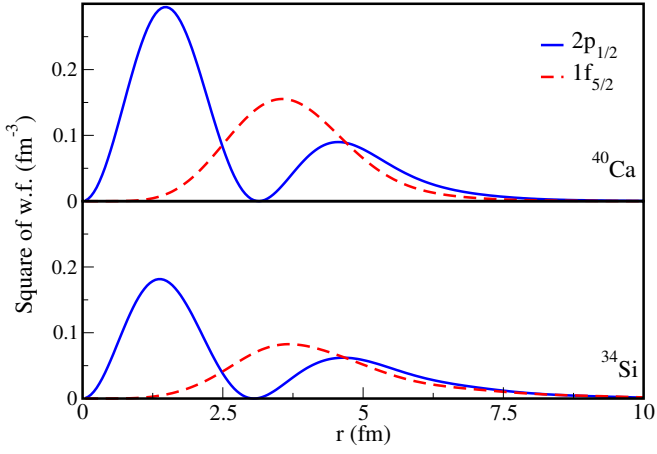}
\caption{ Radial profiles of the $2p_{1/2}$ (full) and the $1f_{5/2}$ (dashed) neutron state for $^{40}$Ca and $^{34}$Si. }
\label{fig2}
\end{figure}

As in the nonrelativistic case~\citep{Grasso2015_PRC92-054316}
the state $1{\it f}_{5/2}$ in the nuclei  $^{40}$Ca, $^{38}$Ar,
and $^{36}$S and the states $1{\it f}_{5/2}$ and $2{\it p}_{1/
2}$ in $^{34}$Si are unbound for all forces we have used. In
contrast to Ref.~\citep{Grasso2015_PRC92-054316} where the
Schr\"odinger equation was diagonalized in a box with finite radius, we expand the single-particle solutions
in an oscillator basis. So instead of
increasing the box radius, we change the number of oscillator
shells. To determine the energies of the unbound states we follow
the same criteria mentioned in Ref.~\citep{Grasso2015_PRC92-054316}. More specifically the energies
of the single-particle resonant states should not change within their
width by changing the number of oscillator shells. Also the radial
profile of those states is similar to that of bound states.
As an example we shown in Fig.~\ref{fig2} the radial
profiles of the wave-functions of the states
$1{\it f}_{5/2}$ and $2{\it p}_{1/2}$ calculated with DD-ME2. For
$^{40}$Ca they are bound and for $^{34}$Si they are unbound.

\subsection{Pure mean-field effects}
%
\begin{table}[t]
\renewcommand\arraystretch{1.5}
\begin{tabularx}{\columnwidth}{@{}lXXXXXXXXX@{}}
\hline
\toprule
\hline
      &  & \multicolumn{2}{c}{$^{40}$Ca}  & \multicolumn{2}{c}{$^{38}$Ar}  & \multicolumn{2}{c}{$^{36}$S}   & \multicolumn{2}{c}{$^{34}$Si}  \\
\hline\midrule\hline
      & $\frac{W_1}{W_2}$  & \multicolumn{1}{c} {\it f}  & \multicolumn{1}{c}  {\it p}  & \multicolumn{1}{c} {\it f}  & \multicolumn{1}{c} {\it p}  & \multicolumn{1}{c} {\it f}  & \multicolumn{1}{c} {\it p}  & \multicolumn{1}{c} {\it f}  & \multicolumn{1}{c} {\it p}  \\
\hline
NL3   & 1.11  & 7.21        & 1.69        & 6.90        & 1.77        & 6.43     & 1.80        & 6.08        & 0.71        \\
NL3$^{*}$  & 1.11     & 7.07        & 1.76        & 6.77        & 1.85        & 6.30     & 1.90        & 5.92        & 0.75        \\
FSUGold & 1.03  & 7.14        & 1.38        & 6.75        & 1.37        & 6.18        & 1.31        & 5.80        & 0.60        \\
DD-ME2 & 1.07 & 7.40        & 1.71        & 7.04        & 1.72        & 6.52        & 1.65        & 6.12        & 0.87        \\
DD-ME$\delta$ & 1.32 & 6.97        & 1.51        & 6.97        & 0.93        & 6.36        & 1.32        & 5.96        & 0.80        \\
DD-PC1 & 1.07 & 7.83        & 1.77        & 7.57        & 1.74        & 7.12        &    1.64     & 6.61        & 0.88        \\
PC-PF1 & 1.11  & 6.88        & 1.76        & 6.64        & 1.87        & 6.25        & 1.93        & 5.87        & 0.84   \\
\hline
   Exp. & & 6.98 & 1.66 &  &  & 5.61 & 1.99 & 5.5 & 1.13 \\
\hline
\bottomrule
\hline
\end{tabularx}
\begin{tabular}{@{}ccccccc@{}}
\hline
\toprule
\hline
 &\multicolumn{2}{c}{$^{40}$Ca $\rightarrow$ $^{36}$S} & \  \  \  \   & \multicolumn{2}{c}{$^{36}$S $\rightarrow$ $^{34}$Si}&
 \\
 \hline
 \midrule
\hline
 & $f$ & $p$ && $f$ & $p$ &  \\
 \hline
   NL3 & 11\%  & -6\% && 5\% & 61\% &\\
   NL3$^{*}$ &  11\% & -8\% && 6\% & 60\%  &\\
   FSUGold & 13\% & 5\% && 6\% & 54\%  &\\
   DD-ME2 & 12\% & 3\% && 6\% & 47\% &\\
   DD-ME$\delta$ & 9\% & 13\% && 6\% & 40\% &\\
   DD-PC1 & 9\% & 8\% && 7\% & 46\% &\\
   PC-PF1 & 9\% & -10\% && 6\% & 57\%&\\
\hline
   Exp. & 20\% & -20\% && 2\% & 43\%&\\
\hline
\bottomrule
\hline
\end{tabular}
\caption{Spin-orbit splittings in MeV (Upper part) and their relative reductions (Lower part) for $f$ and $p$ neutron states in the case of no pairing.}
\label{tab2}
\end{table}

We begin our investigations with simple mean-field calculations without pairing:
we solve the Relativistic Hartree equations and investigate the behavior of the single-neutron energies in the $N=20$ isotone chain. In this case the single-particle orbits are either fully occupied or completely empty. Thus the occupancy of the $2s_{1/2}$ proton state is two for the nuclei $^{40}$Ca, $^{38}$Ar, and  $^{36}$S and zero for $^{34}$Si. This will give us the pure relativistic mean-field effect on the spin-orbit splittings.

The results for this case are given in Table~\ref{tab2}. In the upper part we show the ${\it f} = 1{\it f}_{7/2}-1{\it f}_{5/2}$ and ${\it p} = 2{\it p}_{3/2}-2{\it p}_{1/2}$ energy splittings
for each specific functional and for each of the nuclei $^{40}$Ca, $^{38}$Ar, $^{36}$S, and $^{34}$Si. In the lower part we present the relative reduction of the {\it f} and {\it p} splittings again for every functional, first as we move from $^{40}$Ca to $^{36}$S and then as we go from $^{36}$S to $^{34}$Si. We also show in the last row the experimental values of the splittings and the reductions for $^{40}$Ca~\citep{Uozumi1994_PRC50-263}, $^{36}$S~\citep{Eckle1989_NPA491-205},  and $^{34}$Si~\cite{Burgunder2014_PRL112-042502}.

For $^{40}$Ca we use the values of the centroids for the distribution of the respective fragments. These data can be compared directly with our theoretical results. In the other two cases this is not possible, because the experimental centroids are not known. Therefore for the $2{\it p}_{3/2}-2{\it p}_{1/2}$ in both $^{36}$S and $^{34}$Si we use the major fragment of each state. For the $1f_{5/2}$ state in $^{36}$S we use the major contribution that comes from three states centered at 5.61 MeV with a total spectroscopic factor $SF=0.36$, and in $^{34}$Si  the broad structure around 5.5 MeV with a calculated $SF=0.32$. Even though this is not directly comparable with our results, we use it as an indication of the size of the reduction we should expect.

\begin{figure}[t]
\includegraphics[width=1.0\columnwidth]{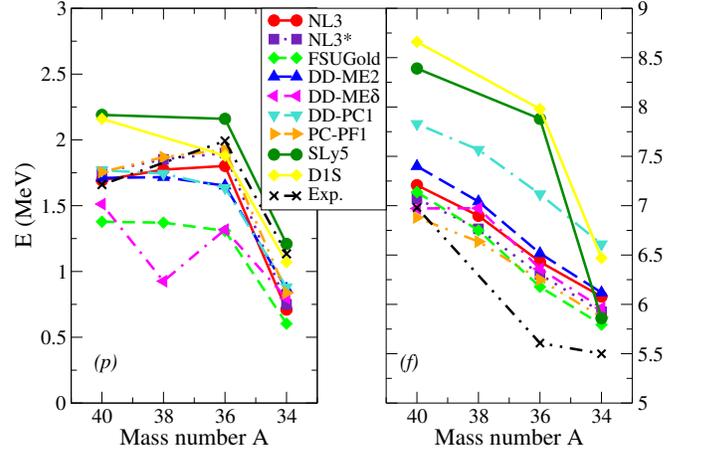}
\caption{(Color online) Evolution of spin-orbit splittings for the neutron levels $p$ (left panel) and $f$ (right panel) with respect to the mass number $A$, without pairing.}
\label{fig3}
\end{figure}
A schematic representation of our results together with the results for the nonrelativistic SLy5 and D1S models, is given in Fig.~\ref{fig3}. For all the models we plot the evolution of the {\it p} and the {\it f} spin-orbit splittings as a function of the mass number $A$.

In this first approach we observe a gradual reduction in the {\it f} splittings of about 0.3-0.4 MeV at each step as we move down the chain of isotones. This is also apparent from the fact that the curves that show the evolution of the \textit{f}-splitting in Fig.~\ref{fig3} have a similar slope for the different functionals. The total relative reduction is between 15\%-19\% and around 5\%-7\% at each step.

In contrast to the {\it f}-splittings, the {\it p} splittings change only slightly for the three first nuclei, the only exception being the functional DD-ME$\delta$. Only when we move from $^{36}$S to $^{34}$Si we find a large reduction for the {\it p} splittings of the order of 40\% to 60\%. Qualitatively this picture is in line with the experiment. However, the absolute size of the \textit{p}-splitting in $^{34}$Si for most of our models is smaller than the respective experimental value. This leads in certain cases to an even larger relative reduction than what we should expect.

The results of the nonrelativistic pure mean-field calculations shown in Table~\ref{tab1} provide a similar qualitative picture. From $^{40}$Ca to $^{36}$S the \textit{f}- and \textit{p}-
splittings are only slightly decreasing with relative reductions 6\%
and 8\% to 8\% and 13\%. In the transition from $^{36}$S to
$^{34}$Si there is also the sudden and relatively large reduction in
the size of the \textit{p}-splitting of about 43\%, but also a bigger
reduction of the size of the \textit{f}-splittings.

When we compare relativistic and nonrelativistic results,
we observe the following differences. In general, the sizes of the
splittings in all the relativistic models are smaller than the
respective splittings in nonrelativistic SLy5 and D1S models. More specifically in
the nuclei $^{40}$Ca and $^{36}$S, where the proton density has the normal profile, i.e., no central depletion, the difference in the size of \textit{f}-splittings is in the order of 1-2 MeV and the size of the \textit{p}-splittings is around 0.5 MeV

In the interesting case of the bubble
nucleus $^{34}$Si, the \textit{f}-splittings are of the same size
because of the bigger relative reduction that appears in the
nonrelativistic case. This is not present in the relativistic models.
However there is a difference in the \textit{p}-splittings which are
relatively small in size for all the relativistic functionals. This is translated into
a relative reduction of the \textit{p}-splitting when we go from $^{36}$S to $^{34}$Si, which
is larger for most of the relativistic models as compared woth the relative reduction for nonrelativistic models
 (see tables~\ref{tab1} and ~\ref{tab2}).

To understand all these results we have to investigate
explicitly the spin-orbit force and especially its isospin dependence
which is very important in the case of $^{34}$Si with
a large neutron-to-proton asymmetry. As we discussed in Sec. \ref{Isospin dependence}, in both
relativistic and nonrelativistic models this force can be approximately written as in
Eq. (\ref{WSO})
\begin{equation}
V_{S.O.}=\bm{W}\cdot \left( \bm{p}\times \bm{\sigma}\right).
\end{equation}%
Here $\bm{W}$ is given by the expression
\begin{equation}
\bm{W}_{\tau }=W_{1}\bm{\nabla}\rho _{\tau }+W_{2}\bm{\nabla}\rho _{\tau
^{\prime }\neq \tau }.
\end{equation}

In most of the nuclei the properties of the nuclear
force lead to an almost constant density in the interior of the
nucleus. The spin-orbit force is mostly determined by the gradient
of the densities and, therefore, by the surface diffuseness. This creates an
attractive potential peaked at the surface. States with large
$\ell$-values have larger $\bm{\ell s}$ values. In addition, they
are peaked near the surface and, therefore, they are influenced more
by this force. This produces the large \textit{f}-splittings and
the much smaller \textit{p}-splittings in $^{40}$Ca, $^{38}$Ar,
and $^{36}$S.

On the other hand, bubble nuclei like
$^{34}$Si have a central density depletion, which provides an additional
component to the spin-orbit force in the interior of the nucleus with the
opposite sign, since the derivative of the density is positive at the origin.
So, together with the attractive well around the surface we also
have a repulsive peak close to the center of the nucleus; see
also Refs.~\citep{Bender1999_PRC60-034304,Todd-Rutel2004_PRC69-021301}.
Neutron states with low angular momentum have larger amplitudes near
the center, as one can see in Fig.~\ref{fig4}. This implies that they
feel a much weaker spin-orbit force and it explains the sudden reduction
of the \textit{p}-splittings when we go from $^{36}$S to $^{34}$Si as shown
in the left panel of Fig.~\ref{fig3}. This effect is not seen
for the \textit{f}-splittings in the relativistic models (right panel of
Fig.~\ref{fig3}).

To understand the aforementioned differences between
relativistic and nonrelativistic models, we concentrate on the
isospin dependence of the SO term $\bm{W}$, which is determined by
the ratio between the two parameters $W_1$ and $W_2$.
In the relativistic models the value of this ratio depends on the
density and can take different values for various nuclei, especially
for functionals where the coupling constants are also density-dependent, as explained in Ref.~\citep{Ebran2016_PRC94-024304}.
In that reference there is a calculation of this ratio for several
nuclei, including $^{34}$Si, as a function of the nuclear radius.
For the functionals DD-ME2 and DD-PC1 at the nuclear
center one has $W1/W2\approx 1.07$. We also give in Table
\ref{tab2} a rough estimate of this ratio for the non-
linear models, using Eq. (\ref{W1/W2_approx}) and neglecting
its density dependence. In general, for the relativistic density functionals,
the value of this ratio is close to unity and the isospin dependence is very weak.
On the other hand, for the standard Skyrme and Gogny models
one has $W1/W2 = 2$ and a stronger isospin dependence.
As it was concluded in Ref.~\citep{Reinhard1995_NPA584-467}, the
additional isospin dependence in the nonrelativistic models
creates a stronger spin-orbit force around the surface and produces
larger splittings for states with large angular momentum.

This picture is reversed in the case of a bubble nucleus, where
the size of the repulsive peak is bigger for the relativistic models, as
very clearly shown in Ref.~\citep{Todd-Rutel2004_PRC69-021301}. As a result
the SO force will be even weaker and the size of the
splitting of the {\it p}-states is more dramatically reduced than in the
standard nonrelativistic forces. Our results lead to the same conclusion.

\subsection{The effect of pairing correlations}
Pairing correlations and the related pairing gap can affect the size of the SO splittings.
Already in Ref.~\cite{Grasso2009_PRC79-034318} it was shown within the framework of relativistic Hartree-Bogoliubov calculations that pairing correlations reduce the size of the bubble in $^{34}$Si. According to this result and based on the previous discussion we expect to see a weakening of the bubble effect and therefore larger absolute sizes and smaller relative reductions of the \textit{p}-splitting, as compared with the pure Hartree-calculations without pairing.

As discussed in Sec. \ref{pairing}, in superfluid nuclei we deal with quasiparticles. The occupancy of each state is calculated self-consistently.
It is determined by the strength of the pairing force. Obviously, for cases with zero pairing the occupation probability is one for occupied states below the Fermi surface and zero for unoccupied states above the Fermi surface. Subsequently, in the present work, we introduce pairing correlations
in the proton subsystem and evaluate again the single-particle energies of the same neutron states as before.
This is done for each nucleus, except from the case of $^{40}$Ca which is a doubly magic nucleus. We also calculate the occupation probabilities of the proton $2s_{1/2}$ state for
$^{36}$S and $^{34}$Si, since the bubble structure in $^{34}$Si is created because of this state being almost empty.

In this context we use the TMR separable pairing force of Ref.~\citep{TIAN-Y2009_PLB676-44}
for the short range correlations. As we mentioned in Sec.~\ref{pairing},
this kind of separable pairing force has been adjusted
to reproduce the pairing gap of the Gogny force D1S in symmetric
nuclear matter~\cite{TIAN-Y2009_PLB676-44}. Both forces are of finite range and therefore they
show no ultraviolet divergence and do not depend on a pairing cut-
off. They provide a very reasonable description of pairing
correlations all over the periodic table with a fixed set of
parameters. However, careful investigations of the size of these
pairing correlations by comparing theoretical results with
experimental odd-even mass differences and experimental
rotational moments of inertia~\cite{Afanasjev2013_PRC88-014320} have shown that the pairing
correlations produced by these forces are slightly too strong for
heavy nuclei and slightly too week for light nuclei. To avoid
such problems in details of the description of pairing correlations in
our relatively light isotonic chain and following the prescription of
Ref.~\cite{Afanasjev2013_PRC88-014320} we have introduced a
scaling factor for the strength of the TMR-force.
To adjust this factor in the proton channel we have used the version of the three-point odd-even staggering (OES) formula proposed in Ref.~\citep{Changizi2015_NPA940-210}:
\begin{equation}{\label{gap formula}}
\Delta^{(3)}_C(N) = \frac{1}{2}[B(N,Z) + B(N - 2,Z) - 2B(N - 1,Z)].
\end{equation}
This is actually equivalent to the original three-point gap formula (\ref{OES3pt-standard}) but given for odd nuclei
$\Delta^3(N-1)$ (see Ref.~\citep{Satula1998_PRL81-3599}).
The binding energies were taken from the atomic mass evaluation in Ref.~\citep{Wapstra2003_NPA729-129} and the resulting gaps are shown in Table~\ref{tab3}.
%
\begin{table}[t]
\renewcommand\arraystretch{1.5}
\begin{tabularx}{0.65\columnwidth}{cXXX}
\hline \toprule \hline
&$^{38}$Ar & $^{36}$S & $^{34}$Si \\
\hline  \hline
$\Delta^{(3)}_C$(MeV)  & 0.93 & 0.45 & 1.95 \\
\hline  \bottomrule \hline
\end{tabularx}
\caption{Gap values calculated with the odd-even mass formula in Eq. (\ref{gap formula}).}
\label{tab3}
\end{table}

\begin{table}[t]
\renewcommand\arraystretch{1.5}
\begin{tabularx}{\columnwidth}{@{}lXXXXXXXX@{}}
\hline
\toprule
\hline
        & \multicolumn{2}{c}{$^{40}$Ca}  & \multicolumn{2}{c}{$^{38}$Ar}  & \multicolumn{2}{c}{$^{36}$S}   & \multicolumn{2}{c}{$^{34}$Si}  \\
\hline\midrule\hline
        & \multicolumn{1}{c} {\it f}  & \multicolumn{1}{c}  {\it p}  & \multicolumn{1}{c} {\it f}  & \multicolumn{1}{c} {\it p}  & \multicolumn{1}{c} {\it f}  & \multicolumn{1}{c} {\it p}  & \multicolumn{1}{c} {\it f}  & \multicolumn{1}{c} {\it p}  \\
\hline
NL3     & 7.21 & 1.69 & 6.92 & 1.64 & 6.46 & 1.68 & 5.94 & 0.80 \\ \midrule
NL3$^{*}$    & 7.07 & 1.76 & 6.78 & 1.76 & 6.32 & 1.80 & 5.77 & 0.85 \\
FSUGold & 7.14 & 1.38 & 6.89 & 1.12 & 6.35 & 1.04 & 5.72 & 0.65 \\
DD-ME2  & 7.40 & 1.71 & 7.08 & 1.64 & 6.55 & 1.57 & 6.00 & 0.94 \\
DD-ME$\delta$  & 6.97 & 1.51 & 6.82 & 1.30 & 6.46 & 1.16 & 5.90 & 0.83 \\
DD-PC1  & 7.83 & 1.77 & 7.58 & 1.67 & 7.14 & 1.56 & 6.52 & 0.96 \\
PC-PF1  & 6.88 & 1.76 & 6.65 & 1.78 & 6.27 & 1.83 & 5.71 & 0.98 \\
\hline
   Exp. & 6.98 & 1.66 &  &  & 5.61 & 1.99 & 5.5 & 1.13\\
\hline
\bottomrule
\hline
\end{tabularx}
\begin{tabular}{@{}ccccccccc@{}}
\hline
 &\multicolumn{2}{c}{$^{40}$Ca $\rightarrow$ $^{36}$S} & \ \ \ \  & \multicolumn{3}{c}{$^{36}$S $\rightarrow$ $^{34}$Si}&
 \\ \midrule
\hline
 & $f$ & $p$ && $f$ & $p$ &  \\
 \hline
   NL3 & 10\%  & 1\% && 8\% & 53\% &\\
   NL3$^{*}$ &  11\% & -3\% && 9\% & 53\% &\\
   FSUGold & 11\% & 24\% && 10\% & 38\% &\\
   DD-ME2 & 11\% & 8\% && 8\% & 40\% &\\
   DD-ME$\delta$ & 7\% & 23\% && 9\% & 28\% &\\
   DD-PC1 & 9\% & 12\% && 9\% & 39\% &\\
   PC-PF1 & 9\% & -4\% && 9\% & 46\% &\\
\hline
   Exp. & 20\% & -20\% && 2\% & 43\% & \\
\hline
\bottomrule
\hline
\end{tabular}
\caption{Same as Table \ref{tab2} but for the case of TMR pairing.}
\label{tab4}
\end{table}
\begin{figure}
\vspace{1cm}
\includegraphics[width=1.0\columnwidth]{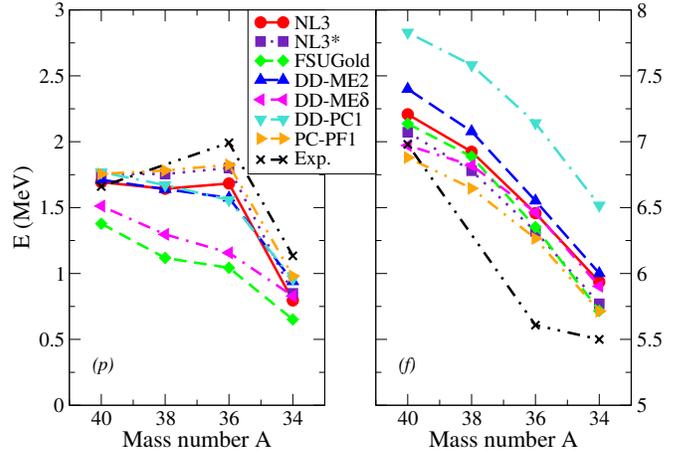}
\caption{ Same as FIG.~\ref{fig3} but with TMR pairing.}
\label{fig4}
\end{figure}

The SO splittings and the respective reductions found in these calculations are shown in Table~\ref{tab4}. In Fig.~\ref{fig4} we present again a schematic representation of the evolution of SO splittings for all the forces with respect to the mass number.

Comparing the results of the calculations including pairing
with the previous pure mean-field results we get the same qualitative picture.
The \textit{f}-splittings show again a gradual reduction as we go
down the chain of isotones. The \textit{p}-splittings stay roughly
in the same size between the first three nuclei and are reduced
dramatically for the last nucleus where there is the bubble
structure.
The inclusion of pairing correlation increases the \textit{f}-
splittings and reduces the \textit{p}-splittings in $^{38}$Ar and
$^{36}$S from the respective splittings in the pure mean-field
calculations. This change is very small for $^{38}$Ar and slightly
bigger for $^{36}$S for the \textit{p}-states and the other way around for the \textit{f}-
splittings, where in the case of $^{36}$S they are practically unchanged.
For the last nucleus $^{34}$Si this picture is reversed and one
gets smaller \textit{f}-splittings and larger \textit{p}-splittings
again in the same order of magnitude of 0.1MeV. This last effect
corrects for the enhanced effect of the bubble structure and the sudden reduction of the \textit{p}-splitting as one goes from $^{36}$S to $^{34}$Si.

%
\begin{table}[t]
\renewcommand\arraystretch{1.5}
\begin{tabular}{@{}ccccccc@{}}
\hline \toprule \hline
 & $^{36}$S  &&$^{34}$Si && $\Delta(2S_{1/2})$ & \\
 \hline \midrule \hline
   NL3 & 1.83  && 0.20 && 1.62 &  \\
   NL3$^{*}$ &  1.87 && 0.23 && 1.64 & \\
   FSUGold & 1.25 && 0.16 && 1.09 & \\
   DD-ME2 & 1.79 && 0.23 && 1.57 &\\
   DD-ME$\delta$ & 1.22 && 0.60 && 1.02 & \\
   DD-PC1 & 1.77 && 0.30 && 1.47 & \\
   PC-PF1 & 1.86 && 0.36 && 1.49 &\\
\hline
   Exp.\cite{Mutschler2016_NatP} & 1.64 && 0.17  && 1.56 & \\
\hline
\bottomrule
\hline
\end{tabular}
\caption{Occupation probabilities of the $2s_{1/2}$ proton state in $^{36}$S and $^{34}$Si for the TMR pairing force.}
\label{tab5}
\end{table}
\begin{figure}
\vspace{1cm}
\includegraphics[width=1.0\columnwidth]{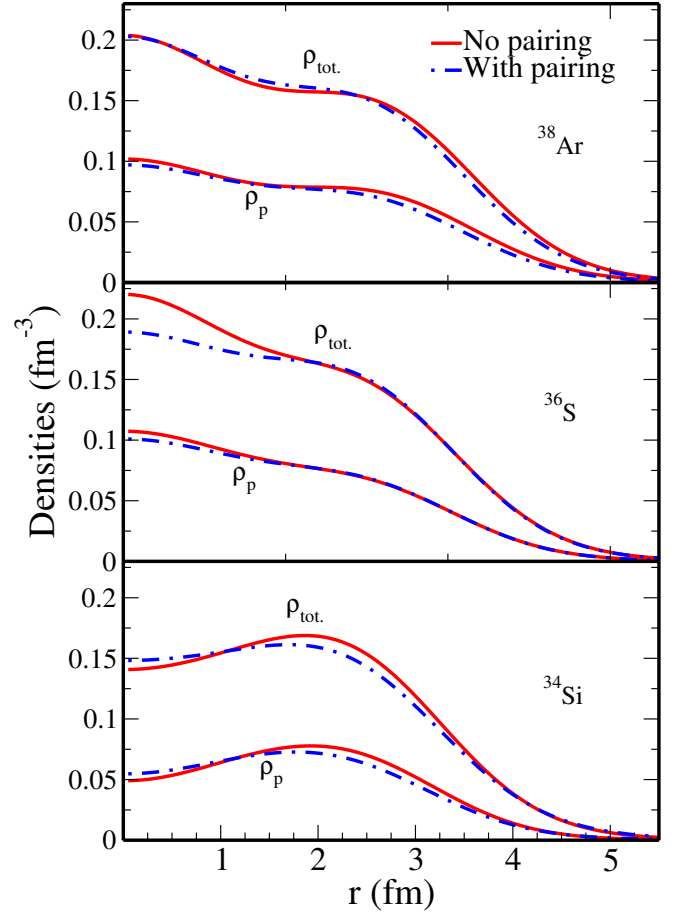}
\caption{ Radial dependence of the total density $\rho_{\rm tot}$ and the proton density $\rho_p$  for NL3 with and without pairing correlations for the nuclei $^{38}$Ar, $^{36}$S, and $^{34}$Si}
\label{fig5}
\end{figure}

For a better understanding how pairing correlations lead to this differences we present in Table \ref{tab5} the occupation factors of the $2s_{1/2}$ proton state in $^{36}$S and $^{34}$Si.
In addition we compare in Fig.~\ref{fig5} the radial profiles of the
total and proton densities of $^{38}$Ar, $^{36}$S, and $^{34}$Si
with and without pairing for the parameter set NL3.

For $^{38}$Ar, pairing affects mostly the \textit{1d} proton orbit with its two last two protons in the $1d_{3/2}$ state. Here the surface density becomes more diffused and the spin-orbit force has a greater overlap with the $f$ neutron states making the corresponding splittings slightly bigger. In the $^{36}$S pairing influences the central densities reducing the size of the peak with a tendency to flatten it out. This can also be seen by the reduced
occupancy of the $2s_{1/2}$ proton state which is now smaller than two. This creates a less attractive SO force around the center and so the splittings of the neutron $p$ states appear somewhat smaller.

For the case of $^{34}$Si pairing reduces the dip at the center of the bubble as it has been noted already in Ref.~\cite{Grasso2009_PRC79-034318}. This is caused by the increasing occupancy of the previously empty $2s_{1/2}$ proton state, as shown in Table \ref{tab5}. As we have seen, this reduction of the bubble leads to an increase of the \textit{p}-splittings by almost 0.1 MeV. Together with the previous discussion about $^{36}$S the relative reduction of this splitting comes closer to the experimental value deduced from the major fragments.
\begin{figure}
\includegraphics[width=1.0\columnwidth]{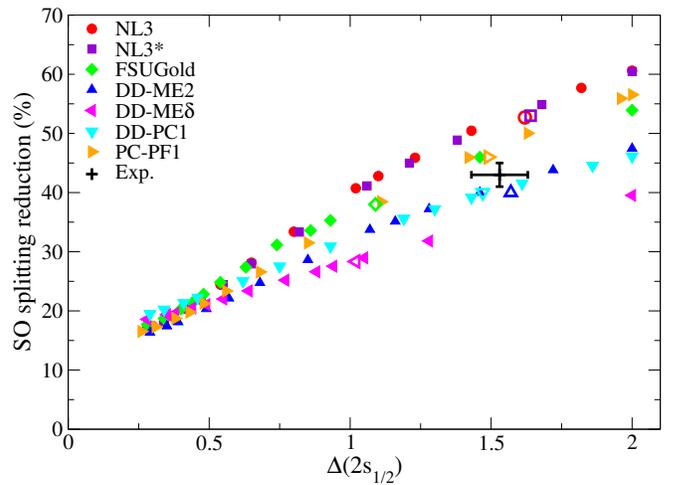}
\caption{Neutron $2p_{1/2} - 2p_{3/2}$ splitting relative
reduction with respect to occupation change in the proton state
$2s_{1/2}$ going from $^{36}$S to $^{34}$Si. More details
are given in the text.}
\label{fig6}
\end{figure}

The above analysis shows that there is a direct relation between the size of \textit{p}-splittings and the occupancy of the $2s_{1/2}$ proton state. In order to elaborate this effect in more detail we carry out RHB-calculations with varying pairing strength by gradually increasing the scaling factor in the TMR-force.
As discussed this leads on one side to a reduction of the corresponding occupancy change $\Delta{}(2s_{1/2})$
between $^{36}$S and $^{34}$Si and on the other side to a reduction of the relative change in the SO-splitting
for the \textit{p}-levels.

As the pairing force increases the bubble structure becomes less dramatic.
Therefore, by studying the corresponding change in the relative reduction of the $2p_{3/2}-2p_{1/2}$ neutron spin-orbit splitting we get an additional method to further investigate the isospin dependence of the effective spin-orbit interaction for the different covariant density functionals. This has been
done in the case of the TMR paring force and for all the relativistic models we have used in our previous calculations and the results are shown in Fig.~\ref{fig6}. The empty symbols depict the results we got using the three-point gap formula to adjust the pairing force. For comparison we show the combined results from the experiments in Refs.~\citep{Burgunder2014_PRL112-042502,Mutschler2016_NatP}.
This helps to distinguish between the various models. We find that DD-ME2, DD-PC1, and PC-PF1 are the most successful in reproducing the experimental results.
\subsection{Extensions: tensor forces and particle-vibration coupling}
In this last part we extend the standard formulation of the
covariant density functional models in two ways. First we include explicitly a tensor term as discussed in Sec.~\ref{tensor-force}. This extension remains on the mean-field level.
In the second case we go beyond mean-field by taking into consideration the coupling of the single-particle states to the low-lying surface modes, as discussed in Sec.~\ref{PVC}.

\subsubsection{The effect of the tensor force}
As we have already stated the tensor part of the nuclear force plays an essential role in the description of the several nuclear properties. In our case it affects the single-particle structure~\cite{Otsuka2005_PRL95-232502,Otsuka2006_PRL97-162501,
Lalazissis2009_PRC80-041301}. As discussed in Sec.~\ref{tensor-force} in covariant density functional theory exchange terms are usually not taken into account, because the Fierz theorem shows that, for zero range forces, they can be expanded over the direct terms by reshuffling the coupling constants of the various spin-isospin channels. Since the coupling constants are adjusted to experimental data anyhow, this seems to be a reasonable approximation for the heavy mesons $\sigma$, $\omega$, and $\rho$, which lead to forces of relatively short range. The direct term of the pion does not contribute because of parity conservation, but its mass is small and, therefore, its exchange term should be taken into account explicitly. It leads to a tensor term in the functional. In the following we show results of relativistic Hartree-Fock calculations as discussed in Sec.~\ref{tensor-force} and in Ref. \cite{Lalazissis2009_PRC80-041301}.

In particular, we investigate in the specific case of the bubble nucleus $^{34}$Si and the corresponding dramatic reduction in the \textit{p}-splitting as compared with $^{36}$S, whether the explicit inclusion of the tensor force changes the size of the splitting and the amount of the reduction.

The effect of the tensor force between neutrons and protons has been investigated in great detail in configuration interaction (CI) calculations~\cite{Otsuka2005_PRL95-232502} and in mean-field calculations~\cite{Otsuka2006_PRL97-162501}. The spin-orbit alignment is very crucial for the attractive or repulsive character of this interaction.
Nucleons occupying, for instance, a proton orbit $j_>$ (where
$j_{>;<}=\ell\pm 1/2$) can change the effective single-particle
energies of neutrons  occupying the orbit $j'_>$ or the orbit $j'_<$
through the monopole effect of the tensor force. If the spins of the
two states are antiparallel, the force is attractive and, if they are
parallel, the force is repulsive. In the particular case of the
one-pion-exchange this specific effect has been also identified in the RHF
calculations of Ref.~\cite{Lalazissis2009_PRC80-041301}.
The effect of the tensor force is mostly important between
neutrons and protons, it increases with the orbital angular
momentum $\ell$ and also with the radial overlap between the orbits.

\begin{figure}[h]
\vspace{0.5cm}
\includegraphics[width=1.0\columnwidth]{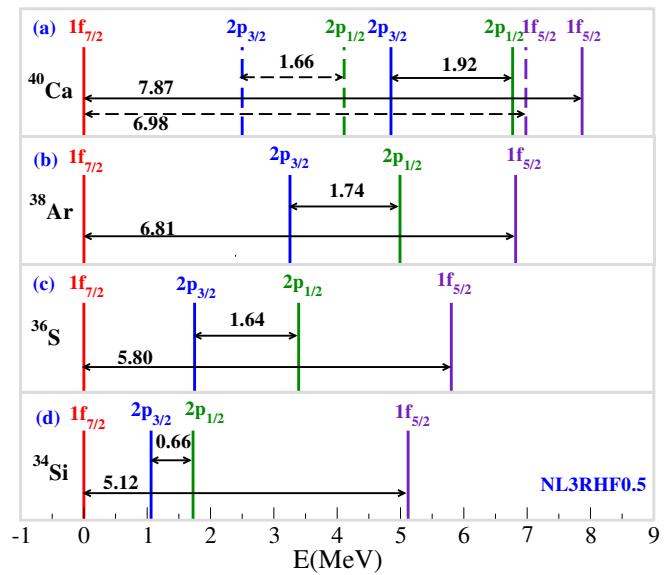}
\caption{The evolution of the energy splittings for the NL3RHF0.5 functional, where boxes (a), (b), (c) and (d) correspond to $^{40}$Ca, $^{38}$Ar, $^{36}$S and $^{34}$Si, respectively. The red dashed lines represent the experimental values of the centroids for $^{40}$Ca~\citep{Uozumi1994_PRC50-263}}
\label{fig7}
\end{figure}

\begin{figure}[!ht]
\includegraphics[width=1.0\columnwidth]{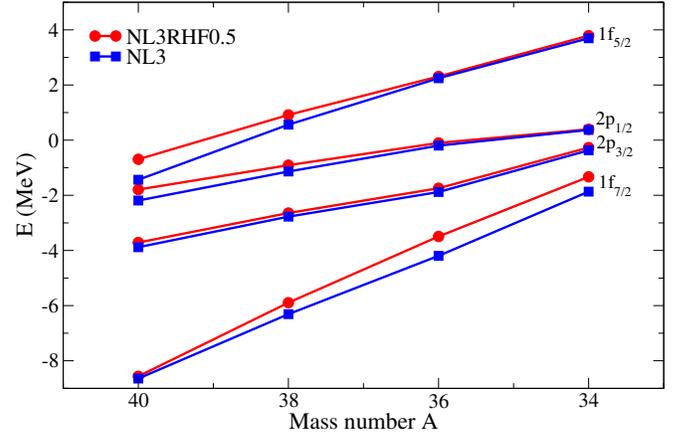}
\caption{Change of single-particle energies of $1f_{5/2}$ and $1f_{7/2}$ and of $2p_{1/2}$ and $2p_3/2$ neutron states as we move down the chain of isotones $N = 20$. The red lines correspond to the tensor results and the blue lines to the standard NL3 with frozen $\Delta$}
\label{fig8}
\end{figure}

\begin{figure}[t]
\includegraphics[width=1.0\columnwidth]{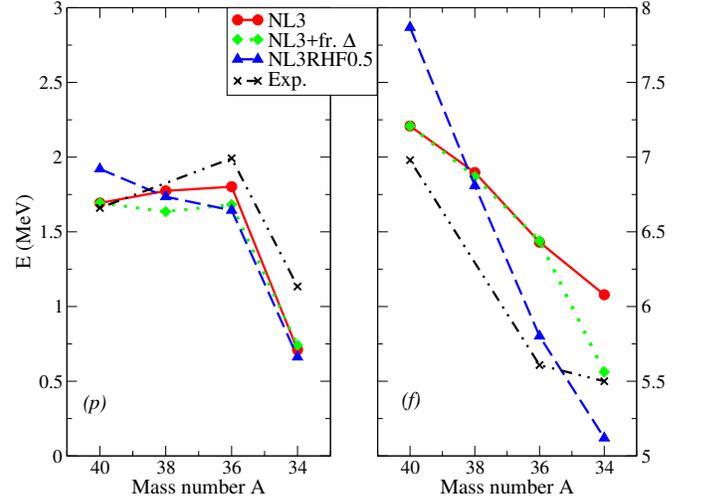}
\caption{Evolution of the {\it p} and {\it f} SO splittings for NL3}
\label{fig9}
\end{figure}

In Fig.~\ref{fig7} we show in a schematic way the positions of the neutron states $2p_{1/2}$, $2p_{3/2}$, and $1f_{5/2}$ using $1f_{7/2}$ as a reference state, calculated with the relativistic
interaction NL3RHF0.5 which includes the one-pion-exchange tensor force with half the strength of the free one-pion-exchange force, as discussed in Sec.~\ref{tensor-force}. These calculations
have been carried out within the frozen gap approximation for the pairing channel and with the values of the proton gap parameters given in Table~\ref{tab3}.
In Table~\ref{tab6} we compare these results with calculation done with
NL3 on the Hartree level with the same pairing scheme of frozen gap and to
those of nonrelativistic Skyrme and Gogny interactions SLy5$_{T-2013}$ and D1ST$_{2c-2013}$.
These are modified versions of the functionals SLy5 and D1S, where tensor
terms have been included and were adjusted together with the spin-orbit
parameters. Details are given in Ref.~\cite{Grasso2015_PRC92-054316}. We have to emphasize, however, that the
tensor force used in the nonrelativistic calculations in Ref.~\cite{Grasso2015_PRC92-054316} is of zero range,
whereas the tensor force in these relativistic calculations is of finite range because of the low mass of the pion.
\begin{table}
\renewcommand\arraystretch{1.5}
\begin{tabularx}{\columnwidth}{@{}ccccccccc@{}}
\hline
\toprule
\hline
        & \multicolumn{2}{c}{$^{40}$Ca}  & \multicolumn{2}{c}{$^{38}$Ar}  & \multicolumn{2}{c}{$^{36}$S}   & \multicolumn{2}{c}{$^{34}$Si}  \\
\hline\midrule\hline
        & \multicolumn{1}{c} {\it f}  & \multicolumn{1}{c}  {\it p}  & \multicolumn{1}{c} {\it f}  & \multicolumn{1}{c} {\it p}  & \multicolumn{1}{c} {\it f}  & \multicolumn{1}{c} {\it p}  & \multicolumn{1}{c} {\it f}  & \multicolumn{1}{c} {\it p}  \\
\hline
NL3    & 7.21 & 1.69 & 6.87 & 1.64 & 6.44 & 1.68 & 5.56 & 0.74 \\
{NL3RHF0.5} & 7.87  & 1.92 & 6.82 & 1.74 & 5.80 & 1.64  & 5.12  & 0.66  \\
{SLy5$_{T-2013}$} & 6.77  & 1.76 &  &  & 5.53  & 1.07  & 4.41
& 0.61  \\
{D1ST$_{2c-2013}$}  & 6.90 & 1.73 & &  & 5.65 & 1.26 & 4.75
& 0.73  \\
 \hline
   Exp. & 6.98 & 1.66 &  &  & 5.61 & 1.99 & 5.5 & 1.13 \\
\hline
\bottomrule
\hline
\end{tabularx}
\begin{tabularx}{0.8\columnwidth}{@{}ccccccc@{}}
\hline\toprule \hline
 &\multicolumn{2}{c}{$^{40}$Ca $\rightarrow$ $^{36}$S} & \ \ \ \  & \multicolumn{2}{c}{$^{36}$S $\rightarrow$ $^{34}$Si}&
 \\ \midrule
\hline
Splitting & $f$ & $p$ && $f$ & $p$ &  \\
 \hline
    NL3 & 10\%  & 1\% && 14\% & 56\% &\\
   {NL3RHF0.5} & 26\%  & 14\% && 12\% & 60\% &\\
   {SLy5$_{T-2013}$} &  18\% & 39\% && 20\% & 43\%  &\\
   {D1ST$_{2c-2013}$} & 18\% & 27\% && 16\% & 42\%  &\\
\hline
   Exp. & 20\% & -20\% && 2\% & 43\%&\\
\hline
\bottomrule
\hline
\end{tabularx}
\caption{Spin-orbit splittings of $f$ and $p$ neutron
states (upper part) and relative reductions (bottom part), for the case of tensor forces. For comparison we also show the results from Ref.~\citep{Grasso2015_PRC92-054316}.}
\label{tab6}
\end{table}
We show in Fig.~\ref{fig8} the corresponding single-particle energies as a function of \textit{A}.
Finally, in Fig.~\ref{fig9} we have plotted the evolution of the spin-orbit splittings as it is done in Figs.~\ref{fig3} and~\ref{fig4}, but now just for the NL3 force in order to compare between pure mean-field, pairing, and tensor effects.

We observe that the inclusion of the tensor force has a more
pronounced effect in the transition from $^{40}$Ca to $^{36}$S
than in the transition from $^{36}$S to $^{34}$Si.
Following the rule that we described in the beginning of the current section,
we recognize that as we move from $^{40}$Ca to $^{36}$S and remove the
four protons from the $j_<$ proton state $\pi 1d_{3/2}$, the attractive effect
of the tensor interaction on the $j'_>$ neutron state $\nu 1f_{7/2}$ is
reduced and, thus, this state is shifted upwards, from its starting point
in $^{40}$Ca. On the other side the $j'_<;\nu 1f_{5/2}$ state, which in $^{40}$Ca
is repelled by the protons of the $\pi 1d_{3/2}$ state, is shifted downwards as we go
to $^{36}$S. The combination of all these effects leads to an
enhanced quenching of the $f$-splitting as we go from $^{40}$Ca to
$^{36}$S. This is also seen by the much steeper blue line that corresponds to NL3RHF0.5 case in the right panel in Fig.~\ref{fig9}.
The same behavior can be observed also for the $j'_>;\nu 2p_{3/2}$ and the $j'_<;\nu 2p_{1/2}$ neutron states, although the effect on the absolute size of the splitting is smaller in those cases.

In the case  of the transition from $^{36}$S to the bubble nucleus $^{34}$Si we see in Fig.~\ref{fig8} that both the $f$ and $p$ states stay at the same distance relative to the NL3 calculations. This shows that the large reduction of the $p$-splitting is a pure spin-orbit effect, a picture that also agrees with the nonrelativistic results.

Finally, we have measured an occupancy of the $2s_{1/2}$ proton state of 0.18 with
the NL3RHF0.5, which is larger than the 0.10 value in the case of
NL3 on the RH level for the same pairing scheme. This indicates that
the tensor force counteracts to some extent the effect of pairing that
we described in the previous section, and leads to a smaller size, from 0.74 MeV to
0.66 MeV and a slightly larger reduction, from 56\% to 60\%, for the particular $p$-splitting.

\subsubsection{The effect of particle-vibration coupling}
As we mentioned in Sec.~\ref{PVC}, the coupling of the single-particle states to low-lying phonons leads to a fragmentation of the single-particle levels and, therefore, sometimes to considerable shifts of the major components, i.e. of the components with the largest spectroscopic factor. This is, in particular, important for states close to the Fermi surface. For our calculations we used the density functional NL3$^{*}$~\cite{Lalazissis2009_PLB671-36} and a constant pairing gap of $\Delta=2$ MeV, which is consistent with
its empirical value of $12.0/\sqrt{A}$ for the considered mass region.

\begin{figure}
\includegraphics[width=1.0\columnwidth]{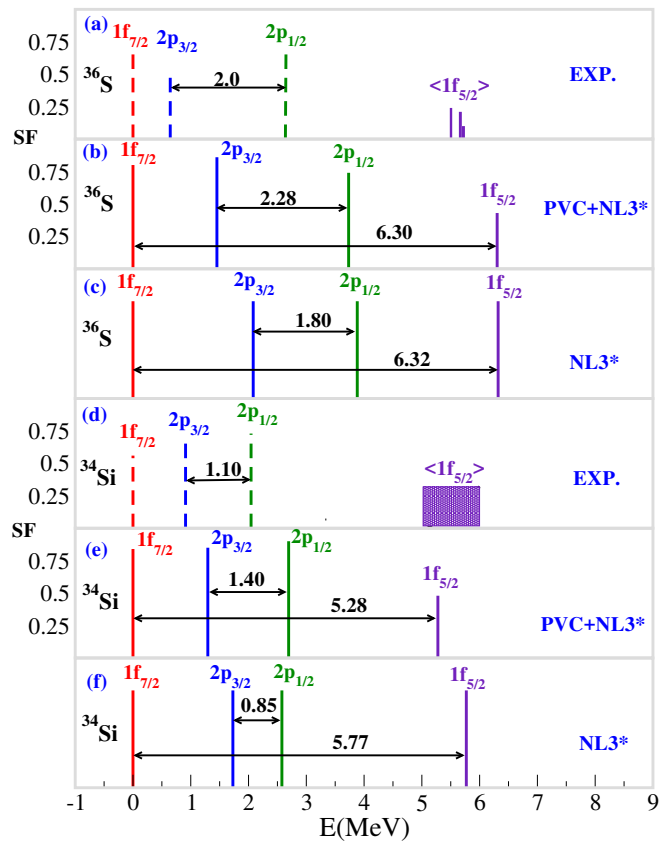}
\caption{Distribution of the major fragments of the single-particle strengths of $^{37}$S (panel (a)) and $^{35}$Si (panel (d)) as given in Ref.~\citep{Burgunder2014_PRL112-042502} and the same distribution calculated with PVC for the force NL3$^{*}$ (panels (b) and (e)). Panels (c) and (f) show results obtained without particle-vibration coupling using the same density functional.}
\label{fig10}
\end{figure}

\begin{table}[ht]
\renewcommand\arraystretch{1.5}
\begin{tabularx}{0.8\columnwidth}{@{}cXXXX@{}}
\hline
\toprule
\hline
        & \multicolumn{2}{c}{$^{36}$S}   & \multicolumn{2}{c}{$^{34}$Si}  \\
\hline\midrule\hline
    Splitting    & \multicolumn{1}{c} {\it f}  & \multicolumn{1}{c} {\it p}  & \multicolumn{1}{c} {\it f}  & \multicolumn{1}{c} {\it p}  \\
\hline
NL3$^{*}$ with PVC    & 6.30  & 2.28 & 5.28  &1.40 \\
Exp.                   & 5.61 & 1.99 & 5.5   & 1.13 \\
\hline
\bottomrule
\hline
\end{tabularx}
\begin{tabular}{@{}cccc@{}}
\hline\toprule \hline
 \ \ \ \  & \multicolumn{2}{c}{$^{36}$S $\rightarrow$ $^{34}$Si}&
 \\ \midrule
\hline
Splitting & $f$ & $p$ &  \\
\hline
 NL3$^{*}$ with PVC  & 16\% & 39\%  &\\
 Exp.                 & 2\% & 43\%  &\\
\hline
\bottomrule
\hline
\end{tabular}
\caption{Comparison for the spin-orbit splittings (Upper part) and their relative reductions (Lower part) of the major fragments between the relativistic PVC model and the corresponding experimental results.}
\label{tab7}
\end{table}
After the solution of the Dyson equation~(\ref{Dyson}) we have the ability to isolate the major contributions to each single particle state and compare its energy directly with the experimental results from Ref.~\citep{Burgunder2014_PRL112-042502}, as shown in table \ref{tab7}. This is also done schematically in Fig.~\ref{fig10} where we compare the results of the PVC calculations
for the nuclei $^{36}$S and $^{34}$Si with the experimental values of Ref.~\citep{Burgunder2014_PRL112-042502}. More specifically, we show the positions of the major fragments and the splittings
between the {\it f} and {\it p} states as well as their
spectroscopic factors. The experimentally observed reduction of the spin-orbit splitting is  $43 \%$ for
the {\it p} states. It is in rather good agreement with
the results obtained from the theoretical PVC calculations, which show a reduction of $39\%$. In both cases
these are the splittings for the major fragments. Notice, that in the PVC calculations we have not included isospin-flip phonons as it is done, for instance, in Ref.~\cite{Litvinova2016_PLB755-138}. It
has been observed that the inclusion of such phonons causes an additional fragmentation and
shifts of the dominant fragments, bringing the results to a better agreement with data. However, the
latter approach is, so far, not yet adopted to the case of open-shell nuclei. It will be considered in the future.

In Fig.~\ref{fig10} we show in analogy to Fig.~\ref{fig7} the positions of the neutron states $2p_{1/2}$, $2p_{3/2}$ and $1f_{5/2}$ using the $1f_{7/2}$ as reference state for the nuclei $^{37}$S [Figs. \ref{fig10}(a)-\ref{fig10}(c)] and $^{35}$Si [Figs. \ref{fig10}(d)-\ref{fig10}(f)]. The experimental data of Ref.~\citep{Burgunder2014_PRL112-042502} in
Figs. \ref{fig10}(a) and \ref{fig10}(d) are compared with results of PVC-calculations with the density functional NL3$^{*}$ in Figs. \ref{fig10}(b) and \ref{fig10}(e). In this figure the experimental energies as well as the energies of the PVC-calculations correspond to the major components of the corresponding fragmented level. Only for the $1f_{5/2}$ orbits we show in Fig. \ref{fig10}(a) the experimental fragmentation and in Fig. \ref{fig10}(d) the area of the experimental fragmentation. In order to study the effect of particle-vibration coupling we show in Figs. \ref{fig10}(c) and \ref{fig10}(f) calculations with the same density functional without particle-vibration coupling.

We find that in both nuclei the SO-splitting of the $1f$ orbitals is reproduced relatively well. Particle vibrational coupling has only a small influence on this splitting. On the other side, all $2p$ orbits are shifted downwards closer to the $1f_{7/2}$ orbit as it is also observed in the experiment. It is well known, that this effect is in particular large for levels close to the Fermi surface, i.e., larger for the $2p_{3/2}$ orbit than for the $2p_{1/2}$ orbit. As a result, the SO-splitting of the $2p$ orbits is increased considerably by particle-vibration coupling. As compared with the much too small SO-splitting for the $2p$ orbits without PVC, it is now much closer to the experimental value.

\section{\label{conclusions}Conclusions}

In this study we have calculated the single-particle energies of the spin-orbit doublets $1f_{7/2}$-$1f_{5/2}$ and $2p_{3/2}$-$2p_{1/2}$ in order to investigate the spin-orbit splittings and their evolution as we move along the chain of isotones with $N=20$: $^{40}$Ca, $^{38}$Ar,  $^{36}$S, and $^{34}$Si. We used several relativistic functionals of three different types: nonlinear meson-coupling, density-dependent meson coupling and density-dependent point-coupling models. Furthermore, we used the separable TMR pairing force of finite range, which is essentially equivalent to the pairing part of the Gogny force D1S, to determine the effect of pairing on the size and on the reduction of the SO splittings.
Finally, we considered specific extensions that go beyond the simple Hartree case; namely, the inclusion of one-pion exchange which induces a tensor force and particle-vibration coupling that takes into account correlations between single-particle states going beyond the mean-field approximation.

In general, we observe a significant reduction of the $2p_{3/2}-2p_{1/2}$ splitting for neutron states when we go from $^{36}$S to $^{34}$Si as it is observed in the experiment. On the pure mean-field level most of the forces show a relatively large reduction. When we include pairing, this reduction becomes less and less dramatic with increasing pairing correlations, because the occupation of the $2s_{1/2}$ proton-orbit changes less rapidly between $^{36}$S and $^{34}$Si. The isospin dependence of the effective spin-orbit force is weaker in the relativistic models and, therefore, the reduction is also less pronounced in these models than in the nonrelativistic ones.

Finally, we went beyond the conventional Hartree level and included two effects, which have a strong influence on the single-particle structure, the tensor term and particle-vibration coupling.

Here we found that the tensor term induced by the one-pion-exchange force has a relatively small effect. It acts to some extent in the opposite direction of pairing. It increases  the quenching of the spin-orbit distinctly for the $f$ and to a smaller extent for the $p$ states when going from $^{40}$Ca to $^{36}$S, showing the tensor character of those reductions.  On the other hand, for the transition from $^{36}$S to $^{34}$Si the sizes of the splittings are only slightly reduced for both nuclei and, thus, the relative reductions remain practically unchanged, indicating that they come purely from the spin-orbit interaction. Such an effect is also observed in the nonrelativistic case~\cite{Grasso2015_PRC92-054316} as seen in Tables \ref{tab1} and \ref{tab6}. However, particle-vibration coupling acts in the same direction as pairing. We find that the relative reduction of the splitting between $2p_{3/2}$ and $2p_{1/2}$ neutron states decreases. This is consistent with the general effect of PVC to produce a more dense spectrum near the Fermi surface. Finally, PVC leads to a reasonable agreement with the experimental data in the isotone chain with $N=20$.

Of course, there are many open questions: Nearly all of the functionals used here have been adjusted to experimental bulk properties, such as binding energies and radii. Only the strength of the relativistic tensor force in the functional NL3RHF0.5 has been optimized at the same time by comparing with the single-particle structure of tin isotopes. In principle, the parameters of all these functionals should be adjusted only after including the additional effects of tensor correlations and particle-vibration coupling. This is a very ambitious task for the future, but we have shown in this investigation at least the influence and the relative importance of several corrections beyond the conventional Hartree level for a successful description of the spin-orbit splitting and its isospin dependence.

\section{acknowledgments}
This research is funded by the Greek State Scholarship Foundation (IKY), through the program \textquotedblleft IKY Fellowships of excellence for postgraduate studies in Greece-SIEMENS program\textquotedblright, by the DFG (Germany) cluster of excellence \textquotedblleft Origin and Structure of the Universe\textquotedblright\ (www.universe-cluster.de), and by NSF (USA) PHY-1404343.


\begin{thebibliography}{10}

\bibitem{Bender2003_RMP75-121}
M. Bender, P.-H. Heenen, and P.-G. Reinhard, Rev. Mod. Phys. {\bf 75},  121
  (2003).

\bibitem{Perdew2004_LNP620-269}
J.~P. Perdew and S. Kurth,  in {\em A Primer in Density Functional Theory},
  edited by C. Fiolhais, F. Nogueira, and M.~A.~L. Marques (Springer Berlin
  Heidelberg, Berlin, 2003), Vol.~620, pp.\ 1--55.

\bibitem{Bartlett2010_MPH108-3299}
R.~J. Bartlett, Mol. Phys. {\bf 108},  3299  (2010).

\bibitem{Vautherin1972_PRC5-626}
D. Vautherin and D.~M. Brink, Phys. Rev. C {\bf 5},  626  (1972).

\bibitem{Decharge1980_PRC21-1568}
J. Decharg\'e and D. Gogny, Phys. Rev. C {\bf 21},  1568  (1980).

\bibitem{Walecka1974_APNY83-491}
J.~D. Walecka, Ann. Phys. (N.Y.) {\bf 83},  491   (1974).

\bibitem{Serot1986_ANP16-1}
B.~D. Serot and J.~D. Walecka, Adv. Nucl. Phys. {\bf 16},  1  (1986).

\bibitem{Boguta1977_NPA292-413}
J. Boguta and A.~R. Bodmer, Nucl. Phys. A {\bf 292},  413  (1977).

\bibitem{Sharma1993_PLB317-9}
M. Sharma, G. Lalazissis, and P. Ring, Phys. Lett. B {\bf 317},  9   (1993).

\bibitem{GoeppertMayer1949_PR75-1969}
M. Goeppert-Mayer, Phys. Rev. {\bf 75},  1969  (1949).

\bibitem{Haxel1949_PR75-1766}
O. Haxel, J.~H.~D. Jensen, and H.~E. Suess, Phys. Rev. {\bf 75},  1766  (1949).

\bibitem{Duerr1956_PR103-469}
H.-P. D{\"u}rr, Phys. Rev. {\bf 103},  469  (1956).

\bibitem{Miller1972_PRC5-241}
L.~D. Miller and A.~E.~S. Green, Phys. Rev. C {\bf 5},  241  (1972).

\bibitem{Baldo2008_PLB663-390}
M. Baldo, P. Schuck, and X. Vi{\~n}as, Phys. Lett. B {\bf 663},  390   (2008).

\bibitem{RocaMaza2011_PRC84-054309}
X. Roca-Maza, X. Vi\~nas, M. Centelles, P. Ring, and P. Schuck, Phys. Rev. C
  {\bf 84},  054309  (2011).

\bibitem{Sharma1994_PRL72-1431}
M.~M. Sharma, G.~A. Lalazissis, W. Hillebrandt, and P. Ring, Phys. Rev. Lett.
  {\bf 72},  1431  (1994).

\bibitem{Reinhard1995_NPA584-467}
P.-G. Reinhard and H. Flocard, Nucl. Phys. A {\bf 584},  467  (1995).

\bibitem{Lalazissis1998_PLB418-7}
G.~A. Lalazissis, D. Vretenar, W. P{\"o}schl, and P. Ring, Phys. Lett. B {\bf
  418},  7  (1998).

\bibitem{Burgunder2014_PRL112-042502}
G. Burgunder, O. Sorlin, F. Nowacki, S. Giron, F. Hammache, M. Moukaddam, N.
  de~S\'er\'eville, D. Beaumel, L. C\`aceres, E. Cl\'ement, G. Duch\^ene, J.~P.
  Ebran, B. Fernandez-Dominguez, F. Flavigny, S. Franchoo, J. Gibelin, A.
  Gillibert, S. Gr\'evy, J. Guillot, A. Lepailleur, I. Matea, A. Matta, L.
  Nalpas, A. Obertelli, T. Otsuka, J. Pancin, A. Poves, R. Raabe, J.~A.
  Scarpaci, I. Stefan, C. Stodel, T. Suzuki, and J.~C. Thomas, Phys. Rev. Lett.
  {\bf 112},  042502  (2014).

\bibitem{Mutschler2016_NatP}
A. Mutschler, A. Lemasson, O. Sorlin, D. Bazin, C. Borcea, R. Borcea, Z.
  Dombradi, J.-P. Ebran, A. Gade, H. Iwasaki, E. Khan, A. Lepailleur, F.
  Recchia, T. Roger, F. Rotaru, D. Sohler, M. Stanoiu, S.~R. Stroberg, J.~A.
  Tostevin, M. Vandebrouck, D. Weisshaar, and K. Wimmer, Nature Phys. {\bf yy},
   xx  (2016).

\bibitem{Grasso2009_PRC79-034318}
M. Grasso, L. Gaudefroy, E. Khan, T. Niksic, J. Piekarewicz, O. Sorlin, N.
  VanGiai, and D. Vretenar, Phys. Rev. C {\bf 79},  034318  (2009).

\bibitem{Uozumi1994_PRC50-263}
Y. Uozumi, N. Kikuzawa, T. Sakae, M. Matoba, K. Kinoshita, T. Sajima, H. Ijiri,
  N. Koori, M. Nakano, and T. Maki, Phys. Rev. C {\bf 50},  263  (1994).

\bibitem{Eckle1989_NPA491-205}
G. Eckle, H. Kader, H. Clement, F. Eckle, F. Merz, R. Hertenberger, H. Maier,
  P. Schiemenz, and G. Graw, Nucl. Phys. A {\bf 491},  205   (1989).

\bibitem{Grasso2015_PRC92-054316}
M. Grasso and M. Anguiano, Phys. Rev. C {\bf 92},  054316  (2015).

\bibitem{Vretenar2005_PR409-101}
D. Vretenar, A.~V. Afanasjev, G.~A. Lalazissis, and P. Ring, Phys. Rep. {\bf
  409},  101  (2005).

\bibitem{VanDalen2007_EPJA31-29}
E.~N.~E. van Dalen, C. Fuchs, and A. Faessler, Eur. Phys. J. A {\bf 31},  29
  (2007).

\bibitem{Lalazissis1997_PRC55-540}
G.~A. Lalazissis, J. K{\"o}nig, and P. Ring, Phys. Rev. C {\bf 55},  540
  (1997).

\bibitem{Lalazissis2009_PLB671-36}
G.~A. Lalazissis, S. Karatzikos, R. Fossion, D. Pe{\~n}a~Arteaga, A.~V.
  Afanasjev, and P. Ring, Phys. Lett. B {\bf 671},  36  (2009).

\bibitem{Todd-Rutel2005_PRL95-122501}
B.~G. Todd-Rutel and J. Piekarewicz, Phys. Rev. Lett. {\bf 95},  122501
  (2005).

\bibitem{Brockmann1992_PRL68-3408}
R. Brockmann and H. Toki, Phys. Rev. Lett. {\bf 68},  3408  (1992).

\bibitem{Typel1999_NPA656-331}
S. Typel and H.~H. Wolter, Nucl. Phys. A {\bf 656},  331   (1999).

\bibitem{Lalazissis2005_PRC71-024312}
G.~A. Lalazissis, T. Nik{\v{s}}i{\'{c}}, D. Vretenar, and P. Ring, Phys. Rev. C
  {\bf 71},  024312  (2005).

\bibitem{Manakos1989_ZPA334-481}
P. Manakos and T. Mannel, Z. Phys. A {\bf 334},  481  (1989).

\bibitem{Nambu1961_PR122-345}
Y. Nambu and G. Jona-Lasinio, Phys. Rev. {\bf 122},  345  (1961).

\bibitem{Burvenich2002_PRC65-044308}
T. B{\"u}rvenich, D.~G. Madland, J.~A. Maruhn, and P.-G. Reinhard, Phys. Rev. C
  {\bf 65},  044308  (2002).

\bibitem{Niksic2008_PRC78-034318}
T. Niksic, D. Vretenar, and P. Ring, Phys. Rev. C {\bf 78},  034318  (2008).

\bibitem{Daoutidis2011_PRC83-044303}
I. Daoutidis and P. Ring, Phys. Rev. C {\bf 83},  044303  (2011).

\bibitem{Koepf1991_ZPA339-81}
W. Koepf and P. Ring, Z. Phys. A {\bf 339},  81  (1991).

\bibitem{Sharma1995_PRL74-3744}
M.~M. Sharma, G. Lalazissis, J. K\"onig, and P. Ring, Phys. Rev. Lett. {\bf
  74},  3744  (1995).

\bibitem{Bulgac1980_nucl-th9907088}
A. Bulgac, Hartree-Fock-Bogoliubov Approximation for Finite Systems, IPNE
  FT-194-1980, Bucharest (arXiv: nucl-th/9907088), 1980.

\bibitem{Dobaczewski1996_PRC53-2809}
J. Dobaczewski, W. Nazarewicz, T.~R. Werner, J.~F. Berger, C.~R. Chinn, and J.
  Decharg\'{e}, Phys. Rev. C {\bf 53},  2809  (1996).

\bibitem{Ring1980}
P. Ring and P. Schuck, {\em The Nuclear Many-Body Problem} (Springer-Verlag,
  Berlin, 1980).

\bibitem{GonzalesLlarena1996_PLB379-13}
T. Gonzalez-Llarena, J. Egido, G. Lalazissis, and P. Ring, Phys. Lett. B {\bf
  379},  13  (1996).

\bibitem{Berger1984_NPA428-23}
J.~F. Berger, M. Girod, and D. Gogny, Nucl. Phys. A {\bf 428},  23c  (1984).

\bibitem{TIAN-Y2009_PLB676-44}
Y. Tian, Z.~Y. Ma, and P. Ring, Phys. Lett. B {\bf 676},  44  (2009).

\bibitem{Berger1991_CPC63-365}
J.~F. Berger, M. Girod, and D. Gogny, Comp. Phys. Comm. {\bf 63},  365
  (1991).

\bibitem{Meng2016_IRNP10}
{\em Relativistic Density Functional for Nuclear Structure}, edited by J. Meng
  (World Scientific, Singapore, 2016), Vol.~10.

\bibitem{Otsuka2005_PRL95-232502}
T. Otsuka, T. Suzuki, R. Fujimoto, H. Grawe, and Y. Akaishi, Phys. Rev. Lett.
  {\bf 95},  232502  (2005).

\bibitem{Brockmann1978_PRC18-1510}
R. Brockmann, Phys. Rev. C. {\bf 18},  1510  (1978).

\bibitem{Bouyssy1987_PRC36-380}
A. Bouyssy, J.~F. Mathiot, N. Van~Giai, and S. Marcos, Phys. Rev. C {\bf 36},
  380  (1987).

\bibitem{Bernardos1993_PRC48-2665}
P. Bernardos, V.~N. Fomenko, N.~V. Giai, M. L. Quelle, S. Marcos, R.
  Niembro, and L.~N. Savushkin, Phys. Rev. C {\bf 48},  2665  (1993).

\bibitem{PhD_Long2005}
W.~H. Long, Phd thesis, Universit{\'e} Paris XI Orsay (unpublished), 2005.

\bibitem{Long2006_PLB640-150}
W.-H. Long, N. Van~Giai, and J. Meng, Phys. Lett. B {\bf 640},  150  (2006).

\bibitem{Long2007_PRC76-034314}
W.-H. Long, H. Sagawa, N. Van Giai, and J. Meng, Phys. Rev. C {\bf 76}, 034314 (2007).

\bibitem{PhD_Serra2001}
M. Serra, Phd thesis, Technical University of Munich (unpublished), 2001.

\bibitem{Lalazissis2009_PRC80-041301}
G.~A. Lalazissis, S. Karatzikos, M. Serra, T. Otsuka, and P. Ring, Phys. Rev. C
  {\bf 80},  041301  (2009).

\bibitem{Schiffer2004_PRL92-162501}
J.~P. Schiffer, S.~J. Freeman, J.~A. Caggiano, C. Deibel, A. Heinz, C.-L.
  Jiang, R. Lewis, A. Parikh, P.~D. Parker, K.~E. Rehm, S. Sinha, and J.~S.
  Thomas, Phys. Rev. Lett. {\bf 92},  162501  (2004).

\bibitem{Litvinova2011_PRC84-014305}
E.~V. Litvinova and A.~V. Afanasjev, Phys. Rev. C {\bf 84},  014305  (2011).

\bibitem{Landau1959_JETP8-70}
L.~D. Landau, Sov. Phys. JETP {\bf 8},  70  (1959).

\bibitem{Migdal1967}
A.~B. Migdal, {\em Theory of Finite Fermi Systems: Applications to Atomic
  Nuclei} (Wilson Interscience, New York, 1967).

\bibitem{Marques2006}
M.~A.~L. Marques, C.~A. Ullrich, F. Nogueira, A. Rubio, K. Burke, and E.~K.
  U.~G. (Eds.), {\em Time-Dependent Density Functional Theory} (Springer,
  Heidelberg, 2006).

\bibitem{Runge1984_PRL52-997}
E. Runge and E.~K.~U. Gross, Phys. Rev. Lett. {\bf 52},  997  (1984).

\bibitem{Litvinova2006_PRC73-044328}
E. Litvinova and P. Ring, Phys. Rev. C {\bf 73},  044328  (2006).

\bibitem{Litvinova2007_PRC75-064308}
E. Litvinova, P. Ring, and V.~I. Tselyaev, Phys. Rev. C {\bf 75},  064308
  (2007).

\bibitem{Litvinova2008_PRC78-014312}
E. Litvinova, P. Ring, and V.~I. Tselyaev, Phys. Rev. C {\bf 78},  014312
  (2008).

\bibitem{Litvinova2012_PRC85-021303}
E. Litvinova, Phys. Rev. C {\bf 85},  021303(R)  (2012).

\bibitem{Gambhir1990_APNY198-132}
Y.~K. Gambhir, P. Ring, and A. Thimet, Ann. Phys. (N.Y.) {\bf 198},  132
  (1990).

\bibitem{Chabanat1998_NPA635-231}
E. Chabanat, P. Bonche, P. Haensel, J. Meyer, and R. Schaeffer, Nucl. Phys. A
  {\bf 635},  231   (1998).

\bibitem{Bender1999_PRC60-034304}
M. Bender, K. Rutz, P.-G. Reinhard, J.~A. Maruhn, and W. Greiner, Phys. Rev. C
  {\bf 60},  034304  (1999).

\bibitem{Todd-Rutel2004_PRC69-021301}
B.~G. Todd-Rutel, J. Piekarewicz, and P.~D. Cottle, Phys. Rev. C {\bf 69},
  021301  (2004).

\bibitem{Ebran2016_PRC94-024304}
J.-P. Ebran, A. Mutschler, E. Khan, and D. Vretenar, Phys. Rev. C {\bf 94},
  024304  (2016).

\bibitem{Afanasjev2013_PRC88-014320}
A.~V. Afanasjev and O. Abdurazakov, Phys. Rev. C {\bf 88},  014320  (2013).

\bibitem{Changizi2015_NPA940-210}
S.~A. Changizi, C. Qi, and R. Wyss, Nucl. Phys. A {\bf 940},  210   (2015).

\bibitem{Satula1998_PRL81-3599}
W. Satu{\l}a, J. Dobaczewski, and W. Nazarewicz, Phys. Rev. Lett. {\bf 81},
  3599  (1998).

\bibitem{Wapstra2003_NPA729-129}
A.~H. Wapstra, G. Audi, and C. Thibault, Nucl. Phys. A {\bf 729},  129
  (2003), the 2003 NUBASE and Atomic Mass Evaluations.

\bibitem{Otsuka2006_PRL97-162501}
T. Otsuka, T. Matsuo, and D. Abe, Phys. Rev. Lett. {\bf 97},  162501  (2006).

\bibitem{Litvinova2016_PLB755-138}
E. Litvinova, Phys. Lett. B {\bf 755},  138   (2016).

\end{thebibliography}

\end{document}